\documentclass[12pt,a4paper]{article}
\usepackage{lscape}
\usepackage{amsfonts}
\usepackage{amsmath}
\usepackage{tabularx}
\usepackage{float}
\usepackage{graphicx,bbm,color}
\usepackage[colorlinks=true,linkcolor=blue]{hyperref}
		\newcommand{\id}{\mathbbm{1}}
		
\usepackage{xcolor,cite,etoolbox}
\makeatletter 
\pretocmd\@bibitem{\color{black}\csname keycolor#1\endcsname}{}{\fail}
\newcommand\citecolor[1]{\@namedef{keycolor#1}{\color{red}}}
\makeatother

\begin{document}
\title{\bf One-dimensional scattering of fermions in double Dirac delta potentials}

\author{L. Santamar\'{\i}a-Sanz$^{1,2}$\footnote{lucia.santamaria@uva.es, lssanz@ubu.es},\\[1ex]
\footnotesize{\sl $^1$Departamento de F\'{\i}sica Te\'{o}rica, At\'{o}mica y \'{O}ptica,
	Universidad de Valladolid, 47011 Valladolid, Spain}\\
	\footnotesize{\sl $^2$Departamento de F\'{\i}sica, Universidad de Burgos, 09001 Burgos, Spain}
}
\date{}
\maketitle
\begin{abstract}
The spectrum of bound and scattering states of the one dimensional Dirac Hamiltonian describing fermions distorted by a static background built from two Dirac delta potentials is studied. A distinction will be made between  `mass-spike' and `electrostatic' $\delta$-potentials. The second quantisation is then performed to promote the relativistic quantum mechanical problem to a relativistic Quantum Field Theory and study the quantum vacuum interaction energy for fermions confined between opaque plates. The work presented here is a continuation of [J.M. Guilarte, J.M. Mu\~noz Casta\~neda, I. Pirozhenko, and L. Santamar\'ia-Sanz. Front. Phys. 7 (2019)].
\end{abstract}

\section{Introduction}

Since its discovery around 1930s, graphene has attracted growing interest because of its applications in Condensed Matter Physics \cite{Wolfbook2013, KatsnelsonSSC2007} and nanoscience \cite{GhoshPRB2008, Milton-PereiraPRB2006}. Graphene consists in a sheet of carbon atoms forming a honeycomb lattice \cite{Wallace1947, NovoselovScience2004}. In the tight-binding approximation \cite{SemenoffPRL1984}, the dynamics of the lower energy charge carriers is described by a $(2+1)$-dimensional massless Dirac-type equation (also called Weyl equation\footnote{For a massive particle, the Dirac wave function is a spinor which describes two spin one-half particles (the particle and its particle). A massless spin one-half particle is described by a Weyl spinor \cite{WeylZPhys1929}.}). Electrons in graphene have an effective velocity 300 times smaller than the velocity of light, allowing the experimental study of interesting relativistic phenomena such as the Klein tunneling \cite{KatsnelsonNature2006}, the Hall efect \cite{GuineaRevModPhys2009, ZhangNature2005} or the Zitterbewegund effect \cite{KatsnelsonEPJB2006}. A wide range of `Dirac materials'\footnote{Namely,  lattice systems where the excitations are described by relativistic Dirac or Weyl equations. These materials are usually narrow gap semiconductors where two or more bands get strongly coupled near a level-crossing. The electrons in these lattices are described by Bloch states. Notice that the aforementioned relativistic equations cannot be satisfied globally over the whole Brillouin zone, but only locally \cite{Cayssol2013}.} like d-wave superconductors \cite{Balatsky2006} and topological insulators \cite{Zhang2007, Zhang2011} share the same fundamental behavior as graphene. They all exhibit universal features such as the same power-law temperature dependence of the fermionic specific heat, the same response to impurities and magnetic fields, the suppression of backscattering, as well as similar transport properties and optical conductivity (see \cite{Wehling2014} for a review). Since these materials can be used to test the predictions of quantum electrodynamics, analysing the dynamics of electrons in them is of great importance both fundamentally and experimentally.

The aim of this work is the study of relativistic Dirac fermionic particles propagating in $(1+1)$-dimensional Minkowski space-time $\mathbb{R}^{1,1}$ while interacting with the static zero range contact potential:
\begin{equation}\label{def-gen-pot}
(q_1 \mathbbm{1}+ \lambda_1 \gamma^0)\, \delta(z+a)+ (q_2 \mathbbm{1}+ \lambda_2 \gamma^0 )\, \delta(z-a), \qquad \lambda_1, \lambda_2, q_1, q_2 \in \mathbb{R},
\end{equation}
with $\gamma^0$ one of the Dirac gamma matrices and $\delta(z)$ the Dirac $\delta$-function. The $\gamma^0$ term  above will contribute to the Dirac Hamiltonian of the theory as a space-dependent mass term, and the one multiplied by $\mathbbm{1}$ as an electromagnetic potential. Fermions propagating on a line are interpreted as the quanta emerging from spinor fields, which are maps from Minkowski space to the fundamental representation of the $\textrm{Spin}(1,1;\mathbb{R})$ group.  The elements of the fundamental representation of $\textrm{Spin}(1,1;\mathbb{R})$ are the spinors \cite{Veltmannbook1994, Friedrichbook2000}, column vectors of two components taking values in the complex field. Studying the spectrum of the fermionic fluctuations interacting with the potential in \eqref{def-gen-pot} is the first step before being able to replicate this potential at a given distance and construct a lattice. In this way, it would be possible to see if there is a connection between the analytical results that can be obtained in the context of Quantum Field Theory (QFT) and the experimental measurements, as well as to study the possible applications in Condensed Matter Physics stated in the first paragraph of this section.

The use of Dirac-type potentials within lattice theories is not a new development. In fact, Dirac delta potentials are widely used as toy models for realistic materials like quantum wires \cite{Cervero2002}, and to analyse physical phenomena such as Bose-Einstein condensation in periodic backgrounds \cite{BordagJPA2020} or light propagation in 1D relativistic dielectric superlattices \cite{Halevi1999}. Despite being a rather simple idealisation of the real system, the $\delta$-function has also been proved to correctly represent surface interactions in many models. For instance, Dirac $\delta$-functions have been set on the plates as models of the electrostatic potential \cite{Hennig1992} or to represent two finite-width mirrors \cite{Fosco2009, PhysRevD.94.125007, PhysRevD.87.025011}. They can also be used to study interactions between semitransparent dielectric surfaces coupled to the electromagnetic field by means of an effective potential\cite{Barone2014}, as well as  to analyse interactions between an atom and a mirror \cite{Oliveira2021}. Furthermore, it is  possible to  describe the permittivity and magnetic permeability in an electromagnetic context by relating the Dirac potential to the plasma frequency in Barton's model on spherical shells \cite{Barton2004, Parashar2012}.  In \cite{GuilarteIJTP2011, GuilarteFrontiers2019} the authors describe how scalar field fluctuations are influenced by this type of singular Dirac potentials. On the other hand, the specific examples of fermions interacting with either a single electric or a mass-spike Dirac delta contact interaction were previously introduced in \cite{GuilarteFrontiers2019}. Now, I want to extend it to double $\delta$-potentials of the form \eqref{def-gen-pot}, before considering the relativistic problem in the associated Dirac combs that can be constructed from this potential.

Once the spectrum of the Dirac Hamiltonian is completely determined, one could focus on the QFT. The use of the theory of self-adjoint extensions of elliptic operators \cite{AIMIJMPA2005} for the computation of the so-called Casimir energy \cite{Miltonbook2001, Bordagbook2009} has motivated an intense research activity so far.  In particular, it has been frequently  used to study scalar quantum fields confined in domains with boundaries (see \cite{AsoreyJPA2006, AsoreyNPB2013, JMMCEPJC2020, BordagEPJC2020} and references therein). Besides the possibility of mimicking impurities in periodic structures, the double $\delta$-potential provides an idealised set up for a pair of partially transparent plates in the Casimir effect \cite{GuilartePRD2013}. Furthermore, the theory of self-adjoint extensions has also been applied for Dirac-type operators \cite{AIMIJMPA2005, AsoreyRVM2016, ElizaldeJPA1998, JaffeAnn2004, DonaireSym2019} to describe fundamental phenomena in topological insulators such as the presence of edge states, as well as to analyse the Casimir pressure for fermionic fields either interacting and confined between plates. My aim will be to continue these studies and generalise them to the case at stake.  I would like to emphasise at this point the importance of studying the spectrum of the problem in quantum mechanics from a mathematical and theoretical point of view before tackling well constructed field theories. To my knowledge, this analysis has not been worked on in such detail before. Consequently, the work presented here lays the groundwork for future projects in QFT with important applications to both high energy and Condensed Matter Physics. Notice that while in the former, the relativistic wave equations in the QFT account for inelastic processes such as particle-antiparticle pair creation, in the latter the Dirac and the Weyl equations can be used directly to describe the band structure of Dirac materials at low energy. In fact, as introduced above, graphene can be studied as a lattice system whose band structure is described by the Weyl equation near some isolated Dirac points of the Brillouin zone in the low-energy limit. At these points, the valence and conducting bands touch. Such Dirac points are protected by space inversion and time-reversal symmetries, and they have been the focus of attention during the last decades since their experimentally observation by the Nobel Prize laureates A. Geim and K. Novoselov in 2005 \cite{NovoselovScience2004,PhysRevB.80.075431}. There exist many experiments which prove the existence of 2D Dirac-Weyl fermions through the measurement of the Quantum Hall effect \cite{ZhangNature2005}. As a consequence, numerous electronic applications such as graphene-based field effect transistors \cite{NovoselovScience2004} have been implemented. Furthermore, it is believed that helical Dirac fermions\footnote{Helical Dirac fermions are charge carriers that behave as massless relativistic particles with an intrinsic angular momentum (spin) locked to its translational momentum. They are guaranteed to be conducting because of time-reversal symmetry, allowing the unique possibility of carrying spin currents without heat dissipation.}, which are forbidden to exist in conventional materials such as graphene or bismuth but may exist at the edges of certain types of topologically ordered insulators, will exhibit topological properties which could manifest as an anomalous half-integer quantisation of Hall conductance, a realisation of Majorana fermions, and a generation of fractionally charged particles \cite{HsiehNature2009}. This will involve the realisation of fundamentally new phenomena in Condensed Matter Physics in the coming years.

So far, only massless fermions have been discussed. However,  massive Dirac fermions are going to be addressed in this work. Notice that it is possible to generate gaps at the Dirac points and to introduce the concept of mass in graphene. One only needs to incorporate a mass term that acts on the sublattice isospin and anticommutes with the Hamiltonian. The simplest choice consists in a constant mass term (independent of the quasi-momentum), similar to that introduced in \eqref{def-gen-pot}. Including this term spoils the inversion symmetry while leaving the time-reversal symmetry intact. The arising gapped electronic system is called Semenov insulator \cite{SemenoffPRL1984} and it is realised for a honeycomb lattice with distinct atoms in the primitive cell sites, such as BN crystals or cold atoms trapped in tunable optical lattices \cite{TarruelNature2012}. Semenov insulators are insulating both in the bulk and along its edges. There are other interesting types of Dirac insulators showing different topological properties. But the most important fact is that recently, it has been discovered that ordinary insulators or semimetallic graphene could be driven into such topological insulators by applying suitable circularly or linearly polarised light \cite{PhysRevB.84.235108, McIver2011}, openning the door to the study of some topological properties and helical fermions. Although these applications are beyond the scope of this paper, it is impossible to end without mentioning that mass fluctuations are of more than just academic importance, and it could be generated in a real experiment via crystal symmetry breaking (by doping techniques and measurement of the detailed resulting band dispersion and density of states of surface states using Landau level spectroscopy at different magnetic fields \cite{OkadaScience2013, ZeljkovicNature2015, WangNature2016}). Seeing the huge progress that is being made in this direction from an experimental point of view, it seems essential to study the theories of massive Dirac fermions interacting with different background potentials from a theoretical and mathematical point of view, as is the main objective of this work.

The article is organised as follows: in section \ref{Sec2} a review of the fundamental concepts of relativistic quantum mechanics is given, and the notation of the article is established. Section \ref{Sec3} collects the description of the spectrum of bound  and scattering states for fermionic fields interacting with a totally generic single Dirac delta potential. Then, in section \ref{Sec4} the same study is carried out for double delta potentials. In section \ref{Sec5}, the second quantisation is performed in order to promote the relativistic quantum mechanical theory presented so far to a relativistic QFT. The transformations of the potential under parity, time reversal and charge conjugation symmetries are summarised. To conclude this work, the study of the quantum vacuum interaction energy for fermions confined between physically opaque plates mimicked by unitary boundary conditions is briefly introduced. Section \ref{Sec6} collects the main conclusions and further work.

\section{Spinor Field Fluctuations}
\label{Sec2}

Throughout the paper I will use natural units $\hbar=c=1$ and consequently,  $M=T=L^{-1}$. Points in the Minkowski space are labelled by real two-vectors $x^\mu\in \mathbb{R}^{1,1}$ with $\mu=0,1$ such that $x^0=t , x^1=z$. Another important vector is the covariant gradient $\partial_\mu\equiv (\partial_0=\partial_t,\partial_1=\partial_z)$. The space-time is equipped with an hyperbolic Lorentzian metric characterised by 
\[
\eta_{\mu\nu}={\rm diag}\{1,-1\}=\eta^{\mu\nu}.
\]
The Clifford algebra associated to the quadratic form defining the metric in $\mathbb{R}^{1,1}$ is given by:
\begin{eqnarray}\label{clif2}
&& \{\gamma^\mu,\gamma^\nu\}=\gamma^\mu\gamma^\nu+\gamma^\nu\gamma^\mu=2\eta^{\mu\nu}, \qquad  \gamma^2= \gamma^0\gamma^1, \qquad \{\gamma^2,\gamma^\mu\}=0,  \qquad  (\gamma^2)^2=\mathbb{I} \, .
\end{eqnarray}
Above, $\gamma^i$ are the Dirac or gamma matrices\footnote{$\gamma^2$ will be the (1+1)-dimensional analogue of  $\gamma^5$ in $\textrm{Cl}\, (\mathbb{R}^{1,3})$.}. I choose the generators of the Clifford algebra as the real $ 2\times 2$ matrices
\begin{equation}
\gamma^0=\sigma_3, \quad \gamma^1=i\sigma_2, \quad  \gamma^2=\sigma_1,\label{gamma-mat}
\end{equation}
where $\{\sigma_1, \sigma_2, \sigma_3\}$ are the Pauli matrices.  The antisymmetric element of the Clifford algebra, $\sigma^{\mu\nu}=i[\gamma^\mu,\gamma^\nu]/4$, generates the uni-parametric ${\rm Spin}(1,1;\mathbb{R})$ Lie group as
\[
S={\rm exp}\left[\frac{i}{2}\omega_{\mu\nu}\sigma^{\mu\nu}\right] \quad \textrm{with}\quad \omega_{01}=-\omega_{10} \in \mathbb{R}.
\] 
The elements of the fundamental representation of this group are the spinors. A minimal representation of the algebra (\ref{clif2}) requires $2\times 2$ matrices so in the fundamental representation, spinors are two-component column vectors: $\Psi=\left(\begin{array}{c}\psi_1\\ \psi_2\end{array}\right)$. The spinor field fluctuations are thus described by maps from the Minkowski space-time to the fundamental representation of ${\rm Spin}(1,1;\mathbb{R})$:
\[
\Psi(x^\mu) : \mathbb{R}^{1,1} \, \longrightarrow \, \mathbb{C} \oplus \mathbb{C} \, .
\]

Furthermore, the most general static backgrounds can be written in the form
\[
V(x^1)=M(x^1)\mathbbm{1}+\gamma^2 \, V_2(x^1)+\gamma^\mu \, V_\mu(x^1),
\]
where $M$ is a Lorentz scalar, $V_2$ one pseudo-scalar and $V_\mu$  one vector potential.  The Einstein summation convention over repeated indices has been used. The influence of $V(x^1)$ on spinor field fluctuations is determined by the Lagrangian densities:
\begin{eqnarray*}
{\cal L}_{\Psi}&=&{\bar \Psi}(x^\mu) \left( i \gamma^\mu \partial_\mu - m-V(x^1)\right)\Psi(x^\mu), \qquad \bar{\Psi}=\Psi^\dagger \gamma^0, \label{lag1}\\
{\cal L}_{\Phi}&=&{\bar \Phi}(x^\mu) \left( i \gamma^\mu \partial_\mu + m-V(x^1)\right)\Phi(x^\mu), \qquad \, \bar{\Phi}=\Phi^\dagger \gamma^0. \label{lag2}
\end{eqnarray*}
 The spinor field equations\footnote{Notice that  in many field theory books the  field equations for the spinor associated to the fermionic particle $\Psi(x^\mu)$ and those relative to the anti-fermionic particle $\Phi(x^\mu)$ are obtained only from one single Lagrangian density. In fact, the Hamiltonians for the free case are related by means of $H_\Phi^{(0)}=-\left(H_\Psi^{(0)}\right)^* $. But here, I am going to treat both problems independently.  Thus the notation used as a subscript in the Lagrangian density and in the successive equations.} read:
\begin{eqnarray}
&& i\partial_0 \Psi(x^0,x^1)=\Big(-i\alpha \partial_1+\beta\left(m+ V(x^1)\right)\Big)\Psi(x^0,x^1), \quad \textrm{with} \quad  \alpha=\gamma^0\gamma^1=\sigma_1, \label{spinfeq1} \\
&& i\partial_0 \Phi(x^0,x^1)=\Big(-i\alpha \partial_1-\beta\left(m- V(x^1)\right)\Big)\Phi(x^0,x^1), \quad \textrm{with} \quad  \beta=\gamma^0=\sigma_3. \label{spinfeq2} 
\end{eqnarray}
The time-energy Fourier transform of the spinor fields, 
\[
\Psi(x^0,x^1)=\int \, \frac{d\omega}{2\pi} \, e^{-i \omega \, x^0} \, \Psi_\omega(x^1) \quad \textrm{and} \quad \Phi(x^0,x^1)=\int \, \frac{d\omega}{2\pi} \, e^{-i \omega \, x^0} \, \Phi_\omega(x^1),
\]
converts the partial differential equations \eqref{spinfeq1}-\eqref{spinfeq2} into the ordinary differential ones:
\begin{eqnarray*}
&& H_\Psi \, \Psi_\omega(z)=\Big(-i \alpha \partial_z +\beta\left(m+V(z)\right)\Big)\Psi_\omega(z)=\omega \Psi_\omega(z), \label{spinfeqs} \\ && H_\Phi \, \Phi_\omega(z)=\Big(-i \alpha \partial_z-\beta\left(m-V(z)\right)\Big)\Phi_\omega(z)=\omega \Phi_\omega(z)\label{spinfeqsc} \, \, .
\end{eqnarray*}
In sum, the problem of determining the spinor field fluctuations in the static background
\begin{equation}
\beta \, V(z)=\gamma^0 \, M(z)+\gamma^1\,  V_2(z)+ \mathbbm{1} \, V_0(z) + \gamma^2\,  V_1(z)  \label{clifpot}
\end{equation}
is equivalent to solve the spectral problem of the Dirac Hamiltonians $H_\Psi$ and $H_\Phi$ in one-dimensional relativistic quantum mechanics. The eigen-spinors of these Hamiltonians will be the one-particle states to be occupied by electrons and positrons moving on a line after the fermionic quantisation procedure be implemented.

Regarding the potential (\ref{clifpot}), since in (1+1)-dimensions there is no magnetic field, a gauge transformation can always be performed so that $V_1(z)=0$ can be taken without loss of generality \cite{JaffeAnn2004}. The original idea of Sundberg and Jaffe is the following: let $\psi(z)$ be a solution of the time-independent Schr\"odinger equation when $V_1(z)=0$. Then the function given by $\varepsilon(z)= A(z) \psi(z)= e^{-i \int_0^z V_1(y)dy}\, \psi(z)$ solves the Schr\"odinger equation for any $V_1(z)$. Notice that $V_1(z)$ only affects the phase of the solutions. The density of states only depends on $z$ and on the energy, but not on $V_1(z)$. Thus, the energy density is independent on $V_1(z)$ and one could consider $V_1(z)=0$ in the model without loss of generality. Furthermore, it is convenient to choose $V_2(z)=0$ because $\gamma^1$ is not Hermitic. Hence, the aforementioned background potential reduces to one term mimicking an electric potential and another one representing a mass term depending on the spatial coordinate $z$. In addition, from now onwards only potentials with compact support are going to be studied:
\begin{equation*}\label{potfer}
V(z)=\left\{ \begin{array}{l l c}
0, &  \textrm{if} & |z|>L,\\
\xi(z)\,  \mathbbm{1} + M(z) \beta,  &  \textrm{if} & -L<z<L.
\end{array}\right.
\end{equation*}In the simplest situation $M(z)=0$, $\xi(z)=0$, described in \cite{GuilarteFrontiers2019}, the eigen-spinors of the Hamiltonians can be easily found.  The basis of states to be used to generate the bound and scattering spinors of the quantum problem in following sections are included in Table \ref{spinorstype}.  
\begin{table}[H]
\centering
\begin{tabularx}{0.95\textwidth} { 
  | >{\centering\arraybackslash}X 
  | >{\centering\arraybackslash}X  
  | >{\centering\arraybackslash}X  | }
 \hline
  \textbf{MOVEMENT}  & \textbf{ELECTRONS  WITH ENERGY $\omega_+>0$}& \textbf{POSITRONS  WITH ENERGY $\omega_+>0$}  \\
 \hline \rule[2mm]{0mm}{4mm}
From left to right with momentum $ k\in \mathbb{R}^+$ &  $\Psi_{+}^{(0)}(t,z) \propto e^{-i \omega_+ t} e^{i k z} \, u_{+}(k)$ &  $\Phi_+^{(0)}(t,z)\propto   e^{-i\omega_+ t}  \, e^{ikz} \, v_+(k) $ \\
\hline \rule[2mm]{0mm}{4mm}
 From right to left  with momentum $-k \in \mathbb{R}^-$  &  $\Psi_{-}^{(0)}(t,z) \propto e^{-i \omega_+ t}\,  e^{-i k z} \,\gamma^0\,  u_{+}(k)$  &  $\Phi_-^{(0)}(t,z) \propto  e^{-i\omega_+ t} \,e^{-ikz} \, \gamma^0\, v_+(k) $\\
\hline
\end{tabularx}
\caption{\footnotesize Positive energy electron spinors versus positron ones.}
\label{spinorstype}
\end{table}
\noindent The basis of eigen-spinors used is given by
\begin{equation*}
u_+(k)=\left(\begin{array}{c}1 \\ \frac{k}{m+\omega_+}\end{array}\right), \qquad v_+(k)=\left(\begin{array}{c}\frac{k}{m+\omega_+} \\ 1\end{array}\right),
\end{equation*}
being $\omega_+=\sqrt{k^2+m^2}$. Notice that the $v_+(k)$ spinors are orthogonal to the positive energy ones $u_+(k)$ because $u^\dagger_+ \gamma^0 v_+=0$.  In sum, both spectra of $H_\Psi^{(0)}$ and $H_\Phi^{(0)}$ are unbounded from below and have a gap $[-m,m]$ with no eigenvalues in between.  Following the Dirac sea prescription\footnote{The infinite Dirac sea proposed by P. Dirac \cite{Dirac1930}  is a theoretical vacuum with only particles with negative energy. The positron was thought as a hole or absence of a particle in the Dirac sea until its discovery as real particle by Carl Anderson \cite{Anderson1933}. Consequently, in the original prescription, spinors associated to electrons with negative energy moving in a certain direction are replaced by spinors of positrons with positive energy moving in the opposite direction.}, only the positive energy eigen-spinors in both problems will be considered. All the negative energy states of both electrons and positrons are filled and the exclusion principle forbids more than one fermionic particle per state. Therefore, positive energy electrons and positrons propagate in $(1+1)$-dimensional Minkowski space-time according to the free Dirac spinors aforementioned.

\section{Dirac spinors interacting with $\delta$-potentials}
\label{Sec3}
Following the notation from Ref. \cite{GuilarteFrontiers2019}, the most general form for a Dirac $\delta$-potential standing at the origin interacting with a Dirac spinor in a $(1+1)$-dimensional space-time reads
\begin{equation}
	V(z)=\Gamma(q,\lambda)\, \delta(z), \quad \textrm{with}\quad \Gamma(q,\lambda)=q\, \mathbbm{1}+\lambda \, \beta.\label{point-V}
\end{equation}
The couplings $\lambda$ and $q$ set the strength of the interactions. Both parameters are dimensionless in the natural system of units. The potential \eqref{point-V} is defined through matching conditions relating the values of the spinors at both sides of the point where the potential stands. As shown in Ref. \cite{GuilarteFrontiers2019}, these matching conditions are given by\footnote{As usual I will use the notation $z=z_0^\pm$ to denote the limit as $z$ approaches to the point $z_0$ from the right ($+$) or from the left ($-$).}
\begin{equation}
\Psi_\omega(0^+)=T_\delta(q,\lambda)\Psi_\omega( 0^-) \, , \, \, \, \quad \Phi_\omega(0^+)=T_\delta(-q,\lambda)\Phi_\omega(0^-)\label{match0},
\end{equation}
where 
\begin{equation}\label{Tmatrixbc}
	T_\delta(q,\lambda)=\cos (\Omega)\mathbbm{1}  -\frac{i}{2}\sin (\Omega) \left[\frac{\Omega}{q+\lambda}(\gamma^2 + \gamma^1)+\frac{q+\lambda}{\Omega}(\gamma^2-\gamma^1)\right],
\end{equation}
and $\Omega\equiv\sqrt{q^2-\lambda^2}$. It is of note that the matching matrix $T_\delta(q,\lambda)$ leaves the potential \eqref{point-V} invariant. In this section, I am going to extend the results publised in \cite{GuilarteFrontiers2019} for a generic single delta potential in which $q, \lambda \neq 0$.
 
\subsection{Scattering states for a single $\delta$-potential.}

Eigen-spinors with $\omega(k)>m>0$ are the scattering states. As happens for the scalar case, there are two independent scattering spinors for a fixed energy. The left-to-right (``{\it diestro}'') spinor for the electrons
\begin{equation}
	\Psi^R_\omega(z,k)=\begin{cases}
		e^{ikz}\,  u_+(k)+\rho _R \, 	e^{-ikz} \, \gamma^0 \, u_+(k),& z<0, \\[1ex]
		\sigma _R\, e^{ikz} \, u_+(k),  & z>0,
	\end{cases}
\label{dscatE}
\end{equation}
and the right-to-left (``{\it zurdo}'') scattering state for electrons
\begin{equation}
	\Psi^L_\omega(z,k)=\begin{cases}
	\sigma _L\, e^{-ikz} \, \gamma^0 \, u_+(k), & z<0, \\[1ex]
		 e^{-ikz} \,\gamma^0\, u_+(k)+\rho _L \,e^{ikz} \,u_+(k), & z>0.
	\end{cases}
	\label{zscatE}
\end{equation}
The scattering amplitudes $\{\sigma_R, \sigma_L,\rho_R, \rho_L\}$ can be obtained imposing the matching conditions \eqref{match0} for the electron spinors above. Solving the two linear systems arising, I obtain the following scattering amplitudes for the electrons on the line interacting with a Dirac $\delta$-potential:
\begin{eqnarray*}
&&\sigma_R (k; \lambda, q)=\sigma_L(k; \lambda, q)=\frac{ k \Omega}{i(q \omega+ m\lambda)\sin \Omega +k \Omega \cos \Omega},\nonumber\\
&&\rho_R(k; \lambda, q)=\rho_L(k; \lambda, q)=\frac{-i \sin \Omega (\omega \lambda + mq) }{i (q \omega + m\lambda)\sin \Omega +k \Omega \cos \Omega}. \label{1deS}
\end{eqnarray*}
It is easy to show that: $\big\vert \sigma(k) \big\vert^2+\big\vert \rho(k)\big\vert^2=1$. In addition, as happens in the scalar case, the reflection and transmission amplitudes are equal for the ``{\it diestro}'' and ``{\it zurdo}'' scattering states. 

The ``{\it diestro}'' and ``{\it zurdo}'' scattering states for the positron spinors can be easily obtained from Eqs. \eqref{dscatE} and \eqref{zscatE} by replacing $u_+(k)$ by $ v_+(k)$ and denoting by $\{\tilde\sigma_R, \tilde\sigma_L,\tilde\rho_R,\tilde \rho_L\}$ the corresponding scattering amplitudes for positrons. Forcing the positron scattering spinors to satisfy the associated matching condition in Eq. \eqref{match0} yields the following scattering amplitudes
\begin{eqnarray*}\label{2deS}
&&\tilde\sigma_R (k; \lambda, q)=\tilde\sigma_L(k; \lambda, q)=\frac{k \Omega }{-i(q\omega + m\lambda )\sin \Omega +  k\Omega \cos \Omega},\nonumber\\
&&\tilde\rho_R(k; \lambda, q)=\tilde\rho_L(k; \lambda, q)=\frac{-i(\omega \lambda + m q)\sin \Omega}{-i(q\omega + m\lambda )\sin \Omega +  k\Omega \cos \Omega}.
\end{eqnarray*} 
As expected, the fact that the ``{\it diestro}'' and ``{\it zurdo}'' scattering amplitudes are equal for the electron (equivalently for the positron) indicates that the Dirac-$\delta$  coupled to relativistic spin-$1/2$ particles is parity and time-reversal invariant. 

\subsection{Bound states for a single $\delta$-potential.}
\begin{figure}[H]
\centering
\includegraphics[width=0.42\textwidth]{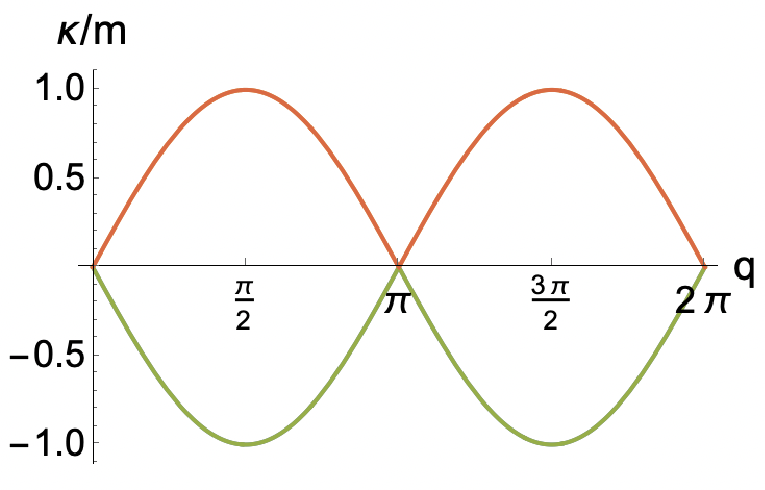}\qquad \qquad \qquad   \includegraphics[width=0.42\textwidth]{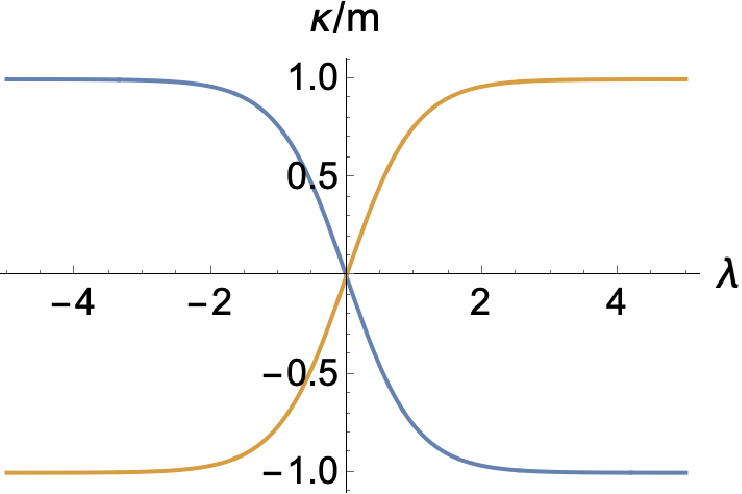}
\caption{\footnotesize Left: Wave vector of bound states in a pure electric $\delta$-potential: $\kappa_{e^-}^+/m$ and $\kappa_{e^+}^+/m$ (green) and $\kappa_{e^-}^-/m$ and $\kappa_{e^+}^-/m$ (red) from Eqs. \eqref{kappael}-\eqref{kappapos} . Right: Wave vector of bound states in a pure massive $\delta$-potential: $\kappa_{e^-}^\pm/m$ (blue) and $\kappa_{e^-}^\pm/m$ (orange) from Eqs. \eqref{kappael}-\eqref{kappapos} . }
\label{kappabs}
\end{figure}

Spinor eigen-functions may also arise if $k=i\kappa$ with $\kappa>0$. Consequently $0<\omega(i\kappa)<m$
and bound states emerge inside the gap when fermions are trapped at the $\delta$-impurity. The ansatz for the bound state spinor wave functions is given by
\begin{equation}
\hspace{-10pt}\Psi^b_\omega(z,\kappa)=\begin{cases}
	A(\kappa)\, e^{\kappa z}\,\gamma^0 \,u_+(i\kappa), & z<0, \\[1ex]
	B(\kappa)\,e^{-\kappa z}\, u_+(i\kappa),  & z>0,
\end{cases}\,\, \, \textrm{and} \,\, \, 
\Phi^b_\omega(z,\kappa)=\begin{cases}
	C(\kappa)\,e^{\kappa z}\,\gamma^0 \,v_+(i\kappa),& z<0, \\[1ex]
	D(\kappa)\,e^{-\kappa z} \,v_+(i\kappa),  & z>0.
\end{cases}
\label{bstE}
\end{equation}
The exponentially decaying solutions of the systems in \eqref{bstE}  (with $\omega^2=m^2-\kappa^2$) in the zone $z<0$ must be related to exponentially decaying solutions of the same systems for $z>0$ by implementing the matching conditions  \eqref{match0} at $z=0$. Doing so yields two linear homogeneous systems. On the one hand, existence of non null solutions for $A$ and $B$ in (\ref{bstE}) requires that the matrix in the corresponding algebraic system has vanishing determinant and consequently:
\begin{equation}
\kappa_{e^-}^\pm= \frac{m \left(\pm 2 q \sqrt{\Omega ^2 (\lambda +q)^4 \sin ^2(\Omega )}+\lambda  \Omega  (\lambda +q)^2 \sin (2 \Omega )\right)}{(\lambda +q)^2 \left(\lambda ^2 \cos (2 \Omega )+\lambda ^2-2 q^2\right)}.\label{kappael}
\end{equation}
On the other hand, non null solutions for $C$ and $D$ in (\ref{bstE}) requires:
\begin{equation}\label{kappapos}
\kappa_{e^+}^\pm=\frac{m \left(\pm 2 q \sqrt{\Omega ^2 (q-\lambda )^4 \sin ^2(\Omega )}-\lambda  \Omega  (q-\lambda )^2 \sin (2 \Omega )\right)}{(q-\lambda )^2 \left(\lambda ^2 \cos (2 \Omega )+\lambda ^2-2 q^2\right)}.
\end{equation}
Figure \ref{kappabs} shows the dependence of $\kappa/m$ with the parameters $q, \lambda$ for a pure electric ($\lambda=0$) delta potential and a pure massive one ($q=0$). Remember that only in the domain of the $q$-$\lambda$ plane where one of the $\kappa$ is real and positive there exists a bound state and one fermion is trapped at the singularity.  

\section{Scattering data and spectrum for double $\delta$ potentials}
\label{Sec4}
The dynamics of fermionic fields interacting with the more general static background that incorporate double $\delta$-potentials symmetrically placed around the origin is governed by the equations
\begin{eqnarray*}
&& i \partial_t \Psi(t,z)= H^{\delta \delta}_\Psi \, \Psi(t,z),\qquad i \partial_t \Phi(t,z)= H^{\delta\delta}_\Phi\,  \Phi(t,z),
\end{eqnarray*}
with the Dirac operators
\begin{eqnarray}
 H^{\delta \delta}_\Psi=-i \alpha\,  \partial_z+ \beta \left[m +\lambda_1\,\delta(z+a)+\lambda_2\,\delta(z-a)\right]+  q_1\,\delta(z+a)+q_2\, \delta(z-a),\label{Hdobleel}
\end{eqnarray}
\vspace{-1cm}
\begin{eqnarray}
H^{\delta\delta}_\Phi =-i \alpha \,\partial_z - \beta [m -\lambda_1\,\delta(z+a)-\lambda_2\,\delta(z-a)]+  q_1\,\delta(z+a)+q_2 \,\delta(z-a)\label{Hdoblepos}.
\end{eqnarray}
Like for single delta potentials, the definition of this background through matching matrices at the singular points $z=\pm a$:
\begin{eqnarray}\label{matchdoble}
&&\hspace{-1.5cm}\left\{ \begin{array}{l}
\Psi_\omega(a^+)= T_\delta(q_2,\lambda_2)\Psi_\omega(a^-)\\
\Psi_\omega(-a^+)= T_\delta(q_1,\lambda_1) \Psi_\omega(-a^-)
\end{array}\right. , \qquad  \qquad  \left\{\begin{array}{l}
\Phi_\omega(a^+)= T_\delta(-q_2,\lambda_2) \Phi_\omega(a^-)\\
\Phi_\omega(-a^+)= T_\delta(-q_1,\lambda_1) \Phi_\omega(-a^-)
\end{array}\right. ,
\end{eqnarray}
is equivalent to provide a self-adjoint extension for the Dirac Hamiltonians.  Two specific examples of double delta potentials will be presented below. 

\subsection{Double electric contact interaction}
Next, only two electric Dirac $\delta$-potentials located at $z=\pm a$ are going to be considered (i.e. the choice $\lambda_1=\lambda_2=0$ is going to be taken in \eqref{Hdobleel}-\eqref{matchdoble}). 

\subsubsection{Electron and positron bound states: the discrete spectrum}
\begin{itemize}
\item \underline{ Electron bound states ($ 0<\kappa < m$)}

\noindent The spinor takes the form
\begin{eqnarray}\label{eq12}
\Psi^b_\omega(z,\kappa)=\left\{
      \begin{array}{ll} 
         A_1(\kappa) \, e^{\kappa z}\,  \gamma^0\,  u_+(i \kappa), & \qquad z< -a,  \\[1ex]
         B_2(\kappa) \, e^{\kappa z}\,  \gamma^0 \, u_+(i \kappa)+ C_2 (\kappa) \, e^{-\kappa z}\,  u_+(i \kappa), & \qquad -a<z<a,  \\[1ex]
         D_3(\kappa) \, e^{-\kappa z}\,  u_+(i \kappa),  & \qquad z> a. 
      \end{array}
   \right.
\end{eqnarray}
The matching conditions  (\ref{matchdoble}) particularised to $\lambda_1=\lambda_2=0$ implies that non trivial solutions for $\{ A_1, B_2, C_2, D_3\}$ exist whenever the following transcendent equation holds:
\begin{equation}
e^{-4 a \kappa}=1+\frac{\kappa [\kappa \cos (q_1+q_2)+ \sqrt{m^2-\kappa^2} \sin(q_1+q_2)]}{m^2 \sin q_1 \sin q_2}. \label{eq35}
\end{equation}
Thus, bound states arise  at the intersections  between the exponential curve $e^{-4 a \kappa}$  and the transcendent one:
\[
Z_1(m,\kappa,q_1,q_2)=1+\frac{\kappa [\kappa \cos (q_1+q_2)+ \sqrt{m^2-\kappa^2} \sin(q_1+q_2)]}{m^2 \sin q_1 \sin q_2},
\]
in the $\kappa\in(0,m)$ open interval, assuming $m>0$. The number of bound states, i.e. the number of intersections between these two curves in the physical range of $\kappa$, depends on the values of the parameters $\{a,m,q_1,q_2\}$. By comparing the tangents of the exponential and the $Z_1$ curves  at  $\kappa=0$, one can see that the identity between them occurs over the curve
\begin{equation}
\frac{\cot q_1 +\cot q_2}{m}=-4 a, \qquad  \rightarrow \qquad \cot q_1 +\cot q_2=-\frac{4}{p}, \label{eq36}
\end{equation}
being $p^{-1}=am$. This trigonometric transcendent equation describes in the $q_1$-$q_2$ plane an ordinary curve which  is a frontier for the number of solutions of (\ref{eq35}) to increase or decrease by one unit.  Furthermore, making $\kappa=m$ in the transcendent spectral equation (\ref{eq35}) yields the condition for the existence of zero modes:
\vspace{-6pt}
\begin{equation}\label{eq37}
e^{-4 a m}=\cot q_1 \cot q_2\qquad  \rightarrow \qquad e^{-4/p}=\cot q_1 \cot q_2.
\end{equation}
The distribution of electron bound sates in the $q_1$-$q_2$ plane is displayed in Figure \ref{fig:mapelel} (left).

It is worth stressing that once $\{q_1,q_2,a,m\}$ take a specific value, the spinor \eqref{eq12} should be normalised. It can be achieved by solving the transcendent equation \eqref{eq35} numerically for these values of $\{q_1,q_2,a,m\}$. Thereafter, one substitutes the specific numerical roots $\kappa$ in the homogeneous linear system resulting from replacing the ansatz of the spinor in the matching conditions \eqref{matchdoble}
to obtain the coefficients $A_1, B_2, C_2, D_3$. Finally, the normalisation condition $|\mathcal{N}|^2 \int_{\mathbb{R}}  \Psi^\dagger(x) \Psi(x) dx=1$ is applied to compute the value of the normalisation constant $\mathcal{N}$.  Figures \ref{fig:groundwave} and \ref{fig:excitedwave} show one specific example of the two bound states spinors arising when $q_1=2, q_2=2.5, a=1, m=1.5$. Notice that to the highest value of $\kappa$ corresponds the lowest bound state energy.

\begin{figure}[H]
\centering
\includegraphics[scale=0.52]{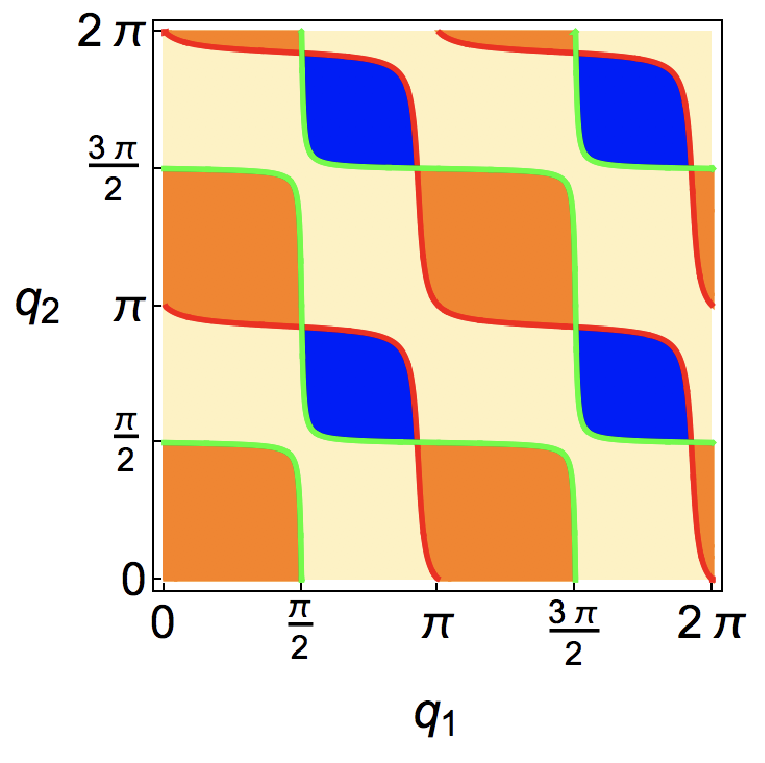} \qquad \qquad \qquad \includegraphics[scale=0.52]{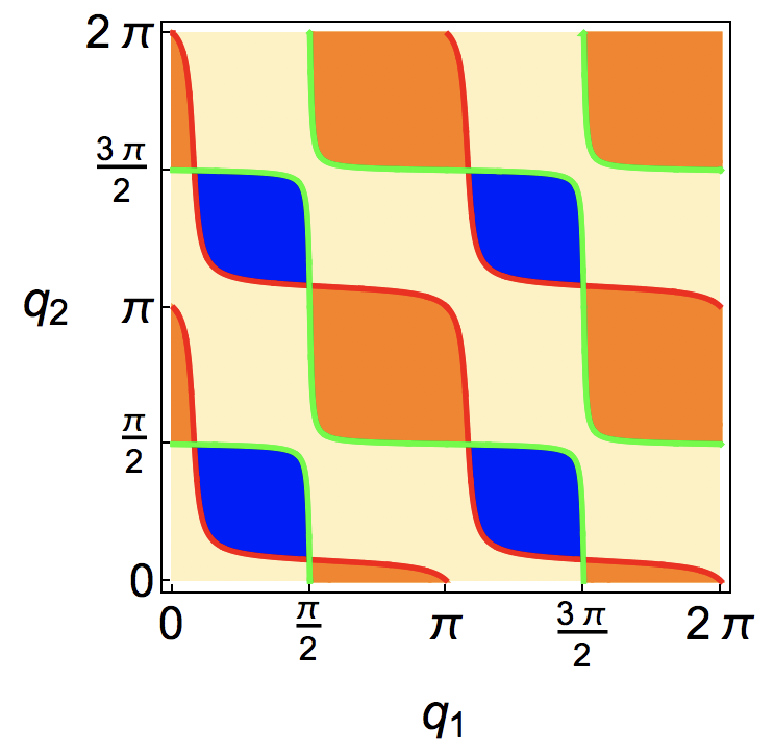}
\caption[\footnotesize   Bound state map in the $q_1,q_2$ space for electrons (left) or positrons (right) in the double electric $\delta$-potential]{\footnotesize   Bound state map in the $q_1,q_2$ space for electrons (left) or positrons (right) trapped by the double electric $\delta$-potential. Blue area: two bound states. Yellow area: one bound state. Orange area: no bound states. The green line characterises the existence of zero modes (\ref{eq37}). The red line is the trigonometric transcendent equation (\ref{eq36}) in the left plot and (\ref{eq40a}) in the right plot. In these plots $a=1, m=1.2$.}
\label{fig:mapelel}
\end{figure}

\begin{figure}[H]
\centering
\includegraphics[scale=0.3]{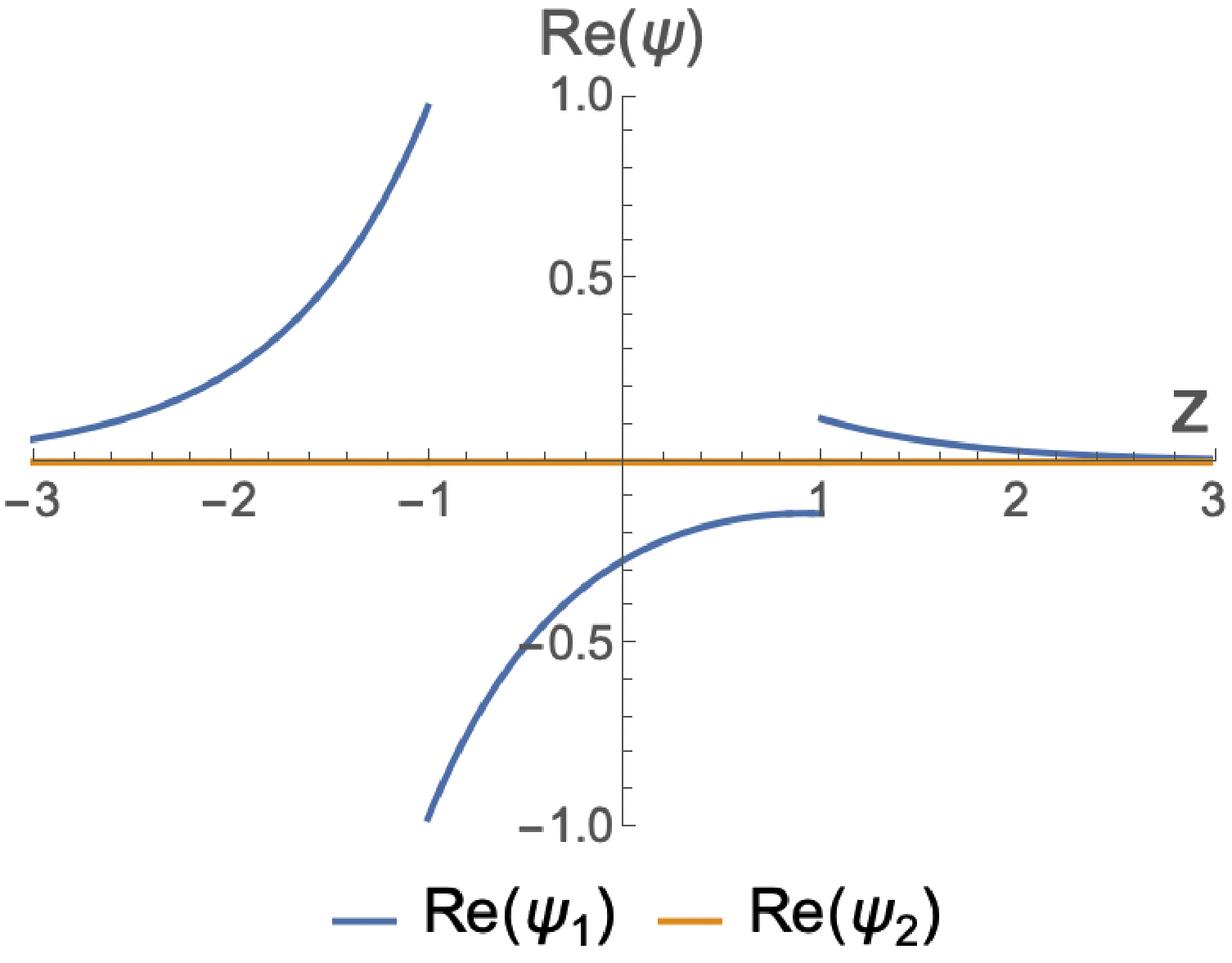} \quad \quad \quad \includegraphics[scale=0.3]{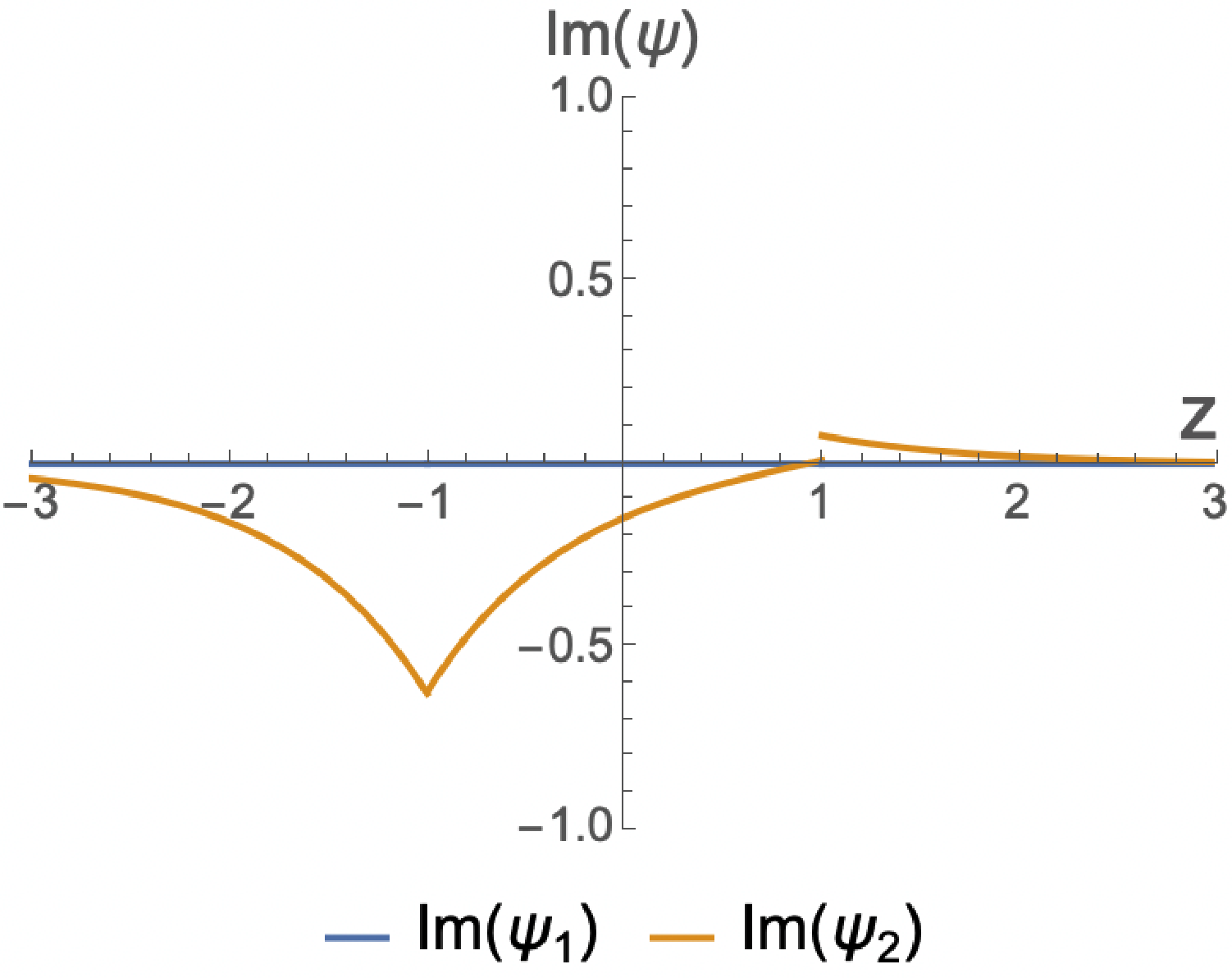}
\caption[\footnotesize  Ground bound state wave function $\Psi^b_{\omega_0}(z, \kappa_0)$ for  $m=1.5, a=1, q_1=2, q_2=2.5$ in a double electric delta interaction]{\footnotesize  Ground bound state wave function $\Psi_{\omega_0}^b(z, \kappa_0)$ for  $m=1.5, a=1, q_1=2, q_2=2.5$.  It corresponds to $\kappa_0=1.3669, \omega_0=0.6177$. Moreover, the numerical coefficients for this example take the following  values: $A_1=1, B_2=-0.0052, C_2=-0.0648, D_3=0.1222, \, \mathcal{N}= \sqrt{14.587}$.}
\label{fig:groundwave}
\end{figure}

\item \underline{ Positron bound state spinors, $ 0<\kappa < m$}

The spinor is the same as \eqref{eq12} but replacing $u_+(i\kappa)$ by $v_+(i\kappa)$. An analogous computation that the one for electrons yields the following transcendent equation:
\begin{equation*}
e^{-4 a \kappa}=1+\frac{\kappa [\kappa \cos (q_1+q_2)- \sqrt{m^2-\kappa^2} \sin(q_1+q_2)]}{m^2 \sin q_1 \sin q_2}. \label{eq39}
\end{equation*}
Now, the trigonometric transcendent equation
\begin{equation}
-\frac{\cot q_1 +\cot q_2}{m}=-4 a, \qquad \rightarrow \qquad  \cot q_1 +\cot q_2=\frac{4}{p},\label{eq40a}
\end{equation}
is the one which describes in the $q_1$-$q_2$ plane an ordinary curve which is a frontier between areas admitting different number of positron bound states. If $\kappa=m$, the positron zero mode existence is also determined by (\ref{eq37}). 
As in the case of electrons, these two curves \eqref{eq37} and \eqref{eq40a} divide the space of parameters  into different zones with zero, one or two bound sates. The distribution of bound states for positrons in the  $q_1$-$q_2$ parameter space is depicted in Figure \ref{fig:mapelel} (right).

\begin{figure}[H]
\centering
\includegraphics[scale=0.3]{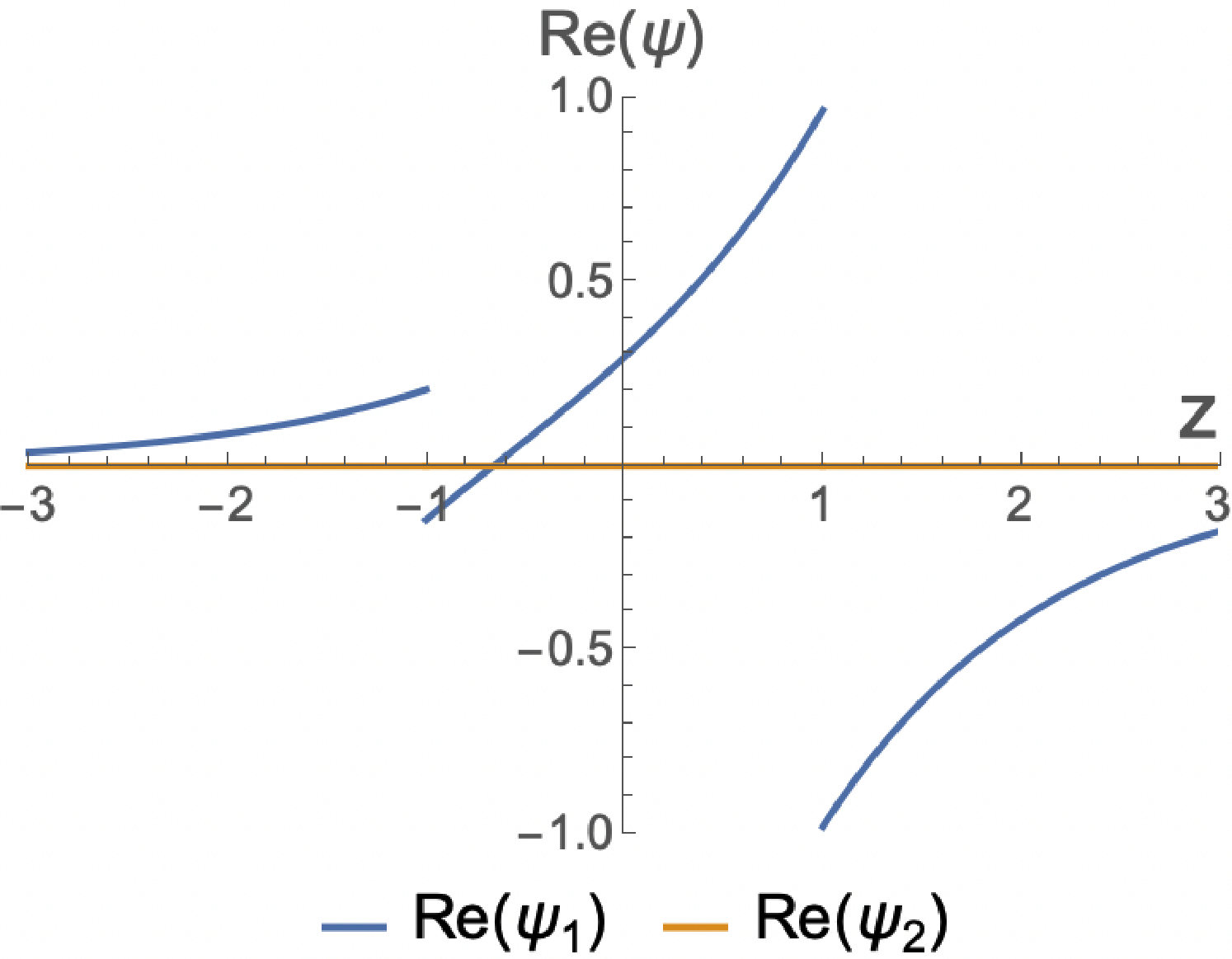} \quad \quad \quad \includegraphics[scale=0.3]{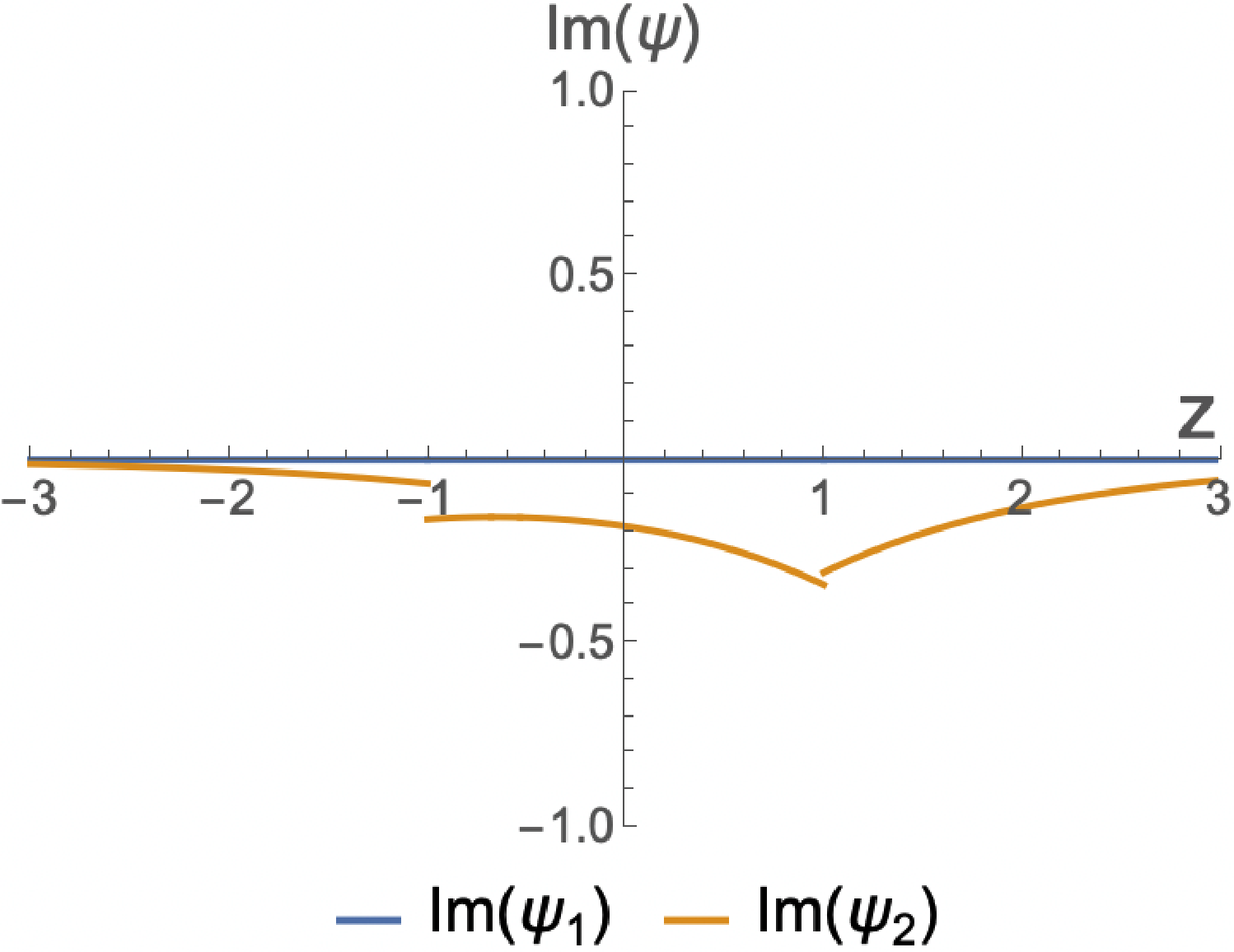}
\caption[\footnotesize   Excited bound state wave function $\Psi^b_{\omega_1}(z, \kappa_1)$ for  $m=1.5, a=1, q_1=2, q_2=2.5$ in a double electric delta interaction]{\footnotesize   Excited bound state wave function $\Psi^b_{\omega_1}(z, \kappa_1)$ for  $m=1.5, a=1, q_1=2, q_2=2.5$.  It corresponds to $\kappa_1=0.8552, \omega_1=1.2323$. Moreover, the numerical coefficients for this example take the following values: $A_1=1, B_2=0.8941, C_2=-0.2883, D_3=-4.7115, \, \mathcal{N}= \sqrt{0.2086}$.}
\label{fig:excitedwave}
\end{figure}

\end{itemize}

Back to the general case,  since there are two electric couplings given by angular coordinates $q_1,q_2 \in [0,2\pi]$ in the model, the space of parameters is the Cartesian product $S^1 \times S^1$ of two circles in $\mathbb{R}^3$. This is a torus, from a topological point of view. Hence, in the natural system of units $\hbar=c=1$, the maximum values of $\{q_1,q_2\}$ together with the mass of the particles and the distance between plates, can be understood as lengths related to the minor and major radius ($r$ and $R$, respectively) of two tori:
\begin{equation*}
T_1\equiv \{R = a \cdot \textrm{max}(q_1), r=\textrm{max}(q_2)/m\},\qquad T_2\equiv \{r = a \cdot \textrm{max}(q_1), R=\textrm{max}(q_2)/m\}.
\end{equation*}

\begin{figure}[H]
\centering
\includegraphics[scale=0.55]{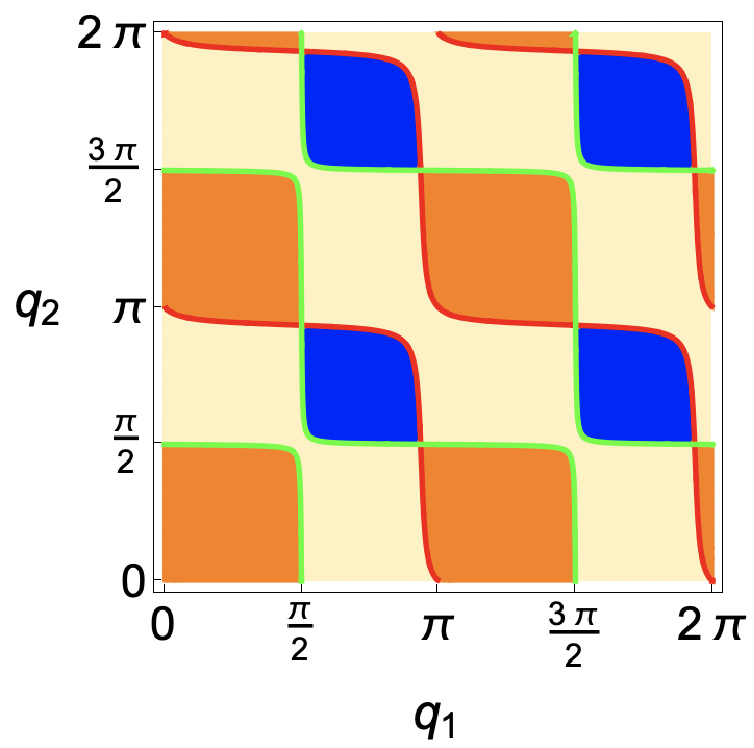} \qquad \qquad \qquad \qquad  \includegraphics[scale=0.4]{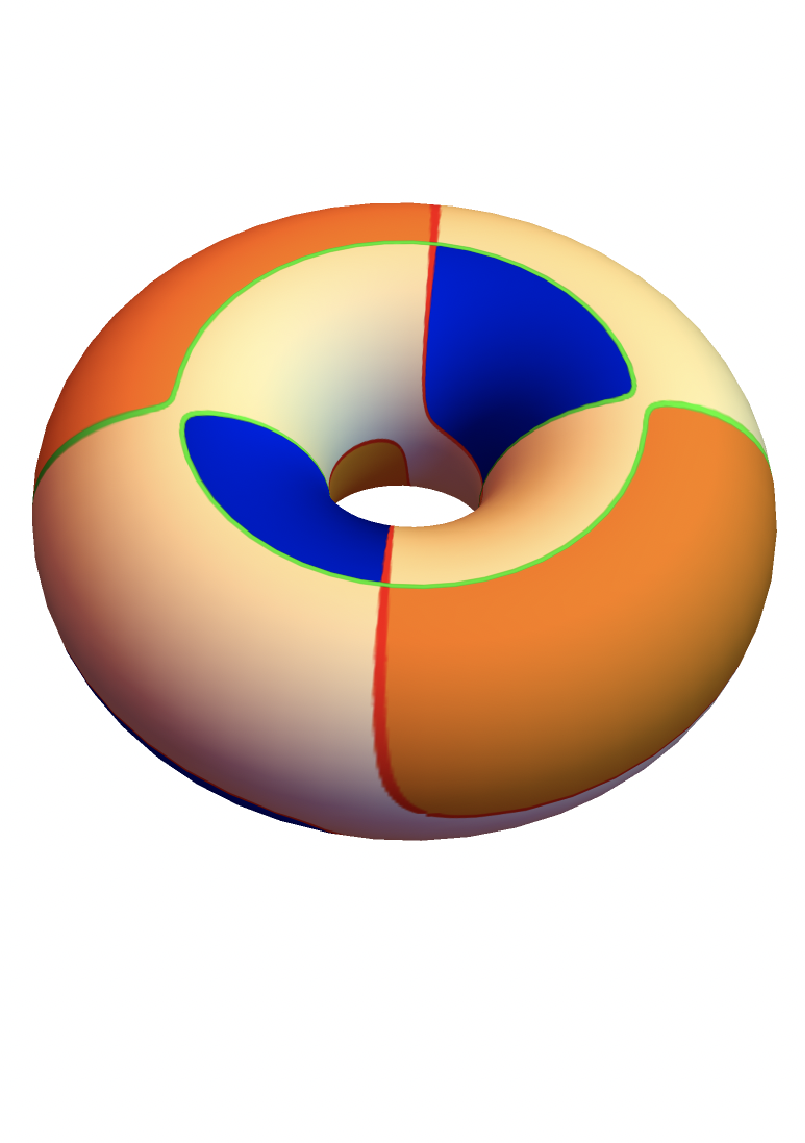} 
\caption{\footnotesize Bound state map for electrons interacting with a double electric delta potential and corresponding complex torus for $p^{-1}=1.5$ with $a=1, m=1.5$.}
\label{fig:torossupercriticos}
\end{figure}

\begin{figure}[H]
\centering
\includegraphics[width=0.93\textwidth]{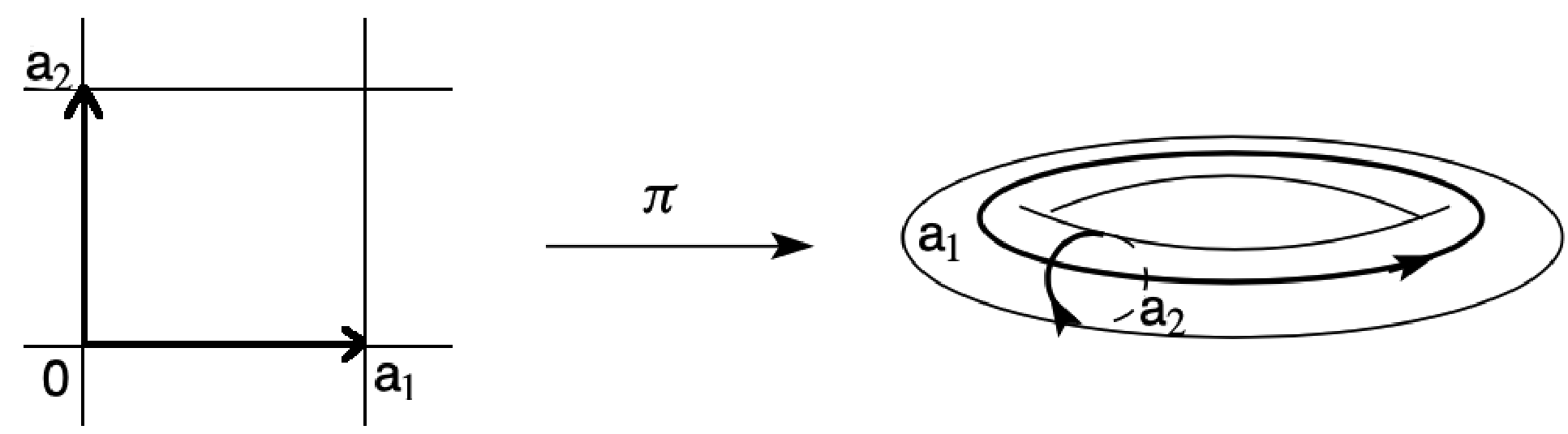} 
\caption[\footnotesize Universal covering map between the lattice generated by $(a_1,a_2)$ and the corresponding torus]{\footnotesize Universal covering map between the lattice generated by $(a_1,a_2)$ and the corresponding torus \cite{Langebook}.}
\label{fig:latticetotorus}
\end{figure}

The transcendent equations \eqref{eq36}, \eqref{eq37} and \eqref{eq40a} describe curves which divide the parameter space into zones with different number of bound states. Furthermore, these curves do not depend on $\{a,m\}$ but on their product. Consequently, one could also represent them over the Riemann surface of the torus as shown in Figure \ref{fig:torossupercriticos}. The parameter $p^{-1}=a\cdot m$ and its inverse fix the complex structure\footnote{ Associating a complex structure means defining the ring of holomorphic and meromorphic functions. A torus may carry a number of different complex structures.} of the two tori associated to the family of theories characterised by $\{a,m\}$.  Notice that here the torus is a connected complex manifold which is homeomorphic to the quotient $\mathbb{C}/L(a_1,a_2)$, being $L(a_1,a_2)$ the lattice generated by $a_1=2\pi a, \, a_2=2\pi/m \in \mathbb{C}$ \cite{Nakaharabook, Langebook}, as seen in Figure \ref{fig:latticetotorus}. Two lattices are equivalent if they are related by the modular group\footnote{The modular group \cite{Gunningbook} is the projective special linear group of $2\times 2$ matrices with integer coefficients and determinant equal to one. Its action on the upper half plan of the complex plane $H$ is the group of linear fractional transformations $$\mathsf{z} \to \frac{a\, \mathsf{z}+b}{c \, \mathsf{z}+d}\, ,$$ with $a,b,c,d \in \mathbb{Z}$ and $ad-bc=1$. The fundamental domain of the modular group can be completely defined as the set $D=\{ \mathsf{z} \in H |\,\,  |\textrm{Re}\, \mathsf{z}|<1/2 \,\, \cup \,\, |\mathsf{z}|>1\}$, whose closure includes at least one point from each equivalence class under the modular group. } $PSL(2,\mathbb{Z})\equiv SL(2, \mathbb{Z})/\mathbb{Z}_2$. The complex structure of a Lie group in the vector space $\mathbb{C}$ induces that of the torus. $\mathbb{C}$ is thus the universal covering space of the torus\footnote{Identifying the opposite sides of the parallelogram gives the torus $T$. Furthermore, there is a universal covering map $\pi: \mathbb{C} \to T$ whose kernel can be identified with the first homology group $\mathcal{H}_1(T, \mathbb{Z})$. Notice that the torus is locally isomorphic to $\mathbb{C}$.}. Hence, the choice of $(a_1,a_2)$ or equivalently $(1, a_2/a_1)$, defines the complex structure of the torus, i.e. the specific way of identifying points in $\mathbb{C}$, modulo $PSL(2,\mathbb{Z})$.

Consequently, when studying fermionic fields in a double electric delta potential, the naturally arising two-parametric family of theories are related to a subset of the moduli\footnote{The moduli is the geometric space where each point represents an isomorphism class of smooth algebraic curves of a fixed genus.} of complex tori or genus one algebraic curves characterised by $ma\in \mathbb{R}$. In such a way, once $\{a, m\}$ are fixed, to each theory corresponds in principle only two tori associated to $p$ and $1/p$. However, one could see that not all $p\in H$ are independent, but the equivalent ones are related by the modular group. Hence, the equivalence class under the modular group $H/PSL(2,\mathbb{Z})$ is the reason that only the torus such that $p^{-1}>1$ (i.e. $a>1/m$) must be taken into account.

\subsubsection{Electron and positron scattering spinors: the continuous spectrum}
\label{subsection4.1.2}
\begin{itemize}

\item \underline{Electron scattering spinor waves: $k\in\mathbb{R}$}.

The scattering spinors for the electrons coming from the left towards the double delta potential (``{\it diestro}'' scattering) are:
\begin{eqnarray}\label{eq22}
\Psi^R_\omega(z,k)=\left\{
      \begin{array}{ll} 
         u_+(k) e^{ikz}+ \rho_R \, \,  \gamma^0\, u_+(k)e^{-ikz},& \qquad z< -a,  \\[1ex]
         A_R \, u_+(k) e^{ikz}+ B_R\,  \, \gamma^0\, u_+(k)e^{-ikz},& \qquad -a<z<a,  \\[1ex]
        \sigma_R  \, u_+(k) e^{ikz}, & \qquad z> a,
      \end{array}
   \right.
\end{eqnarray}
whereas scattering spinors for electrons coming from the right towards the double delta potential (``{\it zurdo}'')  scattering reads
\begin{eqnarray}\label{eq23}
\Psi^L_\omega(z,k)=\left\{
      \begin{array}{ll} 
         \sigma_L\gamma^0\, u_+(k) e^{-ikz},& \qquad z< -a,  \\[1ex]
         A_L \, u_+(k) e^{ik z}+ B_L\,  \, \gamma^0\, u_+(k)e^{-ikz},& \qquad -a<z<a,  \\[1ex]
        \, \gamma^0\, u_+(k)e^{-i k z}+ \rho_L \, u_+(k) e^{i k z}, & \qquad z> a. 
      \end{array}
   \right.
\end{eqnarray}
These piecewise solutions must satisfy the matching conditions (\ref{matchdoble}) for the specific choice $\lambda_1=\lambda_2=0$. Consequently, one obtains two algebraic lineal systems of four equations, one for the four unknowns of the ``{\it diestro}'' scattering $\{\sigma_R,A_R, B_R,\rho_R\}$ and other one for the four unknowns of the ``{\it zurdo}'' scattering $\{\sigma_L,A_L, B_L,\rho_L\}$. Cramer's procedure offers the following solution for the scattering amplitudes:
\begin{eqnarray*}
&& \sigma_R(k;q_1,q_2)=\sigma_L(k;q_1,q_2)=\frac{k^2}{\Lambda(k;q_1,q_2)}=\sigma(k;q_1,q_2), \nonumber\\
&&\rho_R(k;q_1,q_2)=-\frac{2 i m \sqrt{k^2+m^2} \sin q_1 \sin q_2 \, \sin (2ak) + ikm\, \Theta (k; q_1,q_2)}{\Lambda(k;q_1,q_2)},\nonumber\\
&&\rho_L(k;q_1,q_2)=-\frac{2 i m \sqrt{k^2+m^2} \sin q_1 \sin q_2 \, \sin (2ak) + ikm\, \Theta^* (k;q_1,q_2)}{\Lambda(k;q_1,q_2)},\nonumber\\
&&A_L(k; q_1, q_2)=B_R(k; q_1, q_2) = \frac{-i\, k\, m \, e^{2iak}\sin q_1}{\Lambda(k;q_1,q_2)}, \nonumber\\
&&A_R(k; q_1, q_2)=B_L(k; q_1, q_2)=\frac{k^2 \cos q_1 + i k \sqrt{k^2+m^2} \sin q_1}{\Lambda(k;q_1,q_2)},\nonumber\\
&&  \Lambda(k;q_1,q_2)= k^2 \cos (q_1+q_2) +ik \sqrt{k^2+m^2} \sin (q_1+q_2) + m^2 \sin q_1 \sin q_2 (e^{4iak}-1), \nonumber\\
&&\Theta (k;q_1,q_2)=e^{-2iak} \cos q_2 \sin q_1 + e^{2iak} \cos q_1 \sin q_2.
\label{elscatdosel}
\end{eqnarray*}
 It is important to highlight that when only a single delta potential is introduced in the system, the reflection coefficients for diestro and zurdo scattering are equal to each other due to the parity symmetry in the system. Now, when two delta potentials are considered, there could be other type of interactions between the plates due to the quantum vacuum fluctuations  that did not arise in the single $\delta$-case, and parity symmetry could be broken. This is reflected in the fact that now $\rho_L\neq \rho_R$.

\item \underline{Positron scattering spinorial waves: $k\in\mathbb{R}$}. 

The procedure is totally analogous to that of electrons.  The solution of the two algebraic lineal systems of four equations for the  unknowns $\{\tilde{\sigma}_R,\tilde{A}_R, \tilde{B}_R, \tilde{\rho}_R\}$ and for $\{\tilde{\sigma}_L,\tilde{A}_L, \tilde{B}_L, \tilde{\rho}_L\}$ are the following scattering amplitudes:
\begin{eqnarray*}
&& \tilde{\sigma}_R(k;q_1,q_2)=\tilde{\sigma}_L(k;q_1,q_2)=\frac{k^2}{\tilde{\Lambda}(k; q_1,q_2)}=\tilde{\sigma}(k;q_1,q_2),\nonumber\\
&&\tilde{\rho}_R(k;q_1,q_2)=\frac{2i m \sqrt{k^2+m^2} \sin q_1 \sin q_2 \sin(2ak)-ikm\,  \tilde{\Theta} (k; q_1,q_2)}{\tilde{\Lambda}(k;q_1,q_2)},\nonumber\\
&&\tilde{\rho}_L(k;q_1,q_2)=\frac{2 i m \sqrt{k^2+m^2} \sin q_1 \sin q_2 \, \sin (2ak) -ikm\, \tilde{\Theta}^* (k; q_1,q_2)}{\tilde{\Lambda}(k;q_1,q_2)},\nonumber\\
&&\tilde{A}_L(k;q_1,q_2) =\tilde{B}_R(k;q_1,q_2) = \frac{-i\, k\, m \,e^{2iak} \, \sin q_1}{\tilde{\Lambda}(k;q_1,q_2)}, \nonumber\\
&&\tilde{B}_L(k;q_1,q_2) = \tilde{A}_R(k;q_1,q_2) = \frac{k^2 \cos q_1 -ik \sqrt{k^2+m^2} \sin q_1}{\tilde{\Lambda}(k;q_1,q_2)},\nonumber\\
&& \tilde{\Lambda}(k;q_1,q_2)= k^2 \cos (q_1+q_2) -ik \sqrt{k^2+m^2} \sin (q_1+q_2) + m^2 \sin q_1 \sin q_2 (e^{4iak}-1),\nonumber\\
&&\tilde{\Theta} (k; q_1,q_2) =e^{-2iak} \cos q_2 \sin q_1 + e^{2iak} \cos q_1 \sin q_2.
\label{posscatdosel}
\end{eqnarray*}
\end{itemize}
\vspace{0.3cm}

\subsection{Double mass spike contact interaction}
Consider finally the one-dimensional Hamiltonians for the fermionic particle and antiparticle moving in the real line in the background of two mass-like Dirac $\delta$-potentials centred at $z=\pm a$, i.e. \eqref{Hdobleel} and \eqref{Hdoblepos} for $q_1=q_2=0$. Now, the spectral problems for the Dirac Hamiltonian and its conjugate must be solved including the matching conditions defined in (\ref{matchdoble}) for $q_1=q_2=0$.

\subsubsection{Electron and positron bound states: the discrete spectrum}
\begin{itemize}

\item \underline{ Electron bound state spinors, $ 0<\kappa < m$.}

Pugging the  ansatz \eqref{eq12} in the matching conditions (\ref{matchdoble}) for $q_1=q_2=0$, defines a homogeneous linear system in the unknowns $(A_1, B_2, C_2, D_3)$ with non trivial solutions providing:
\begin{equation}
e^{-4 a \kappa}=\frac{(m+\kappa \coth \lambda_1)(m+\kappa \coth \lambda_2)}{m^2-\kappa^2}. \label{specc}
\end{equation}
The number of bound states depends again on the parameters $\{m, a,\lambda_1,\lambda_2\}$ through
\begin{equation}
\coth \lambda_1 +\coth \lambda_2=-\frac{4}{p}, \qquad p^{-1}=am. \label{idbtan}
\end{equation}
The shape of the curve that this hyperbolic transcendent equation describes in the $\lambda_1$-$\lambda_2$ plane is similar to that of a hyperbola with two branches. One of the vertices is placed at the origin, the other at the point $(\lambda_1=\lambda_2=-{\rm arccoth}\, 2)$, and the axis is the $\lambda_1=\lambda_2$ straight line. For points above the upper branch of the curve, no bound states are encountered. Points in between the two branches correspond to one bound state. Points in the zone below the lower branch give rise to two bound states. This distribution can be seen in Figure \ref{fig:mapmasel} (left).

\begin{figure}[H]
\centering
\includegraphics[scale=0.32]{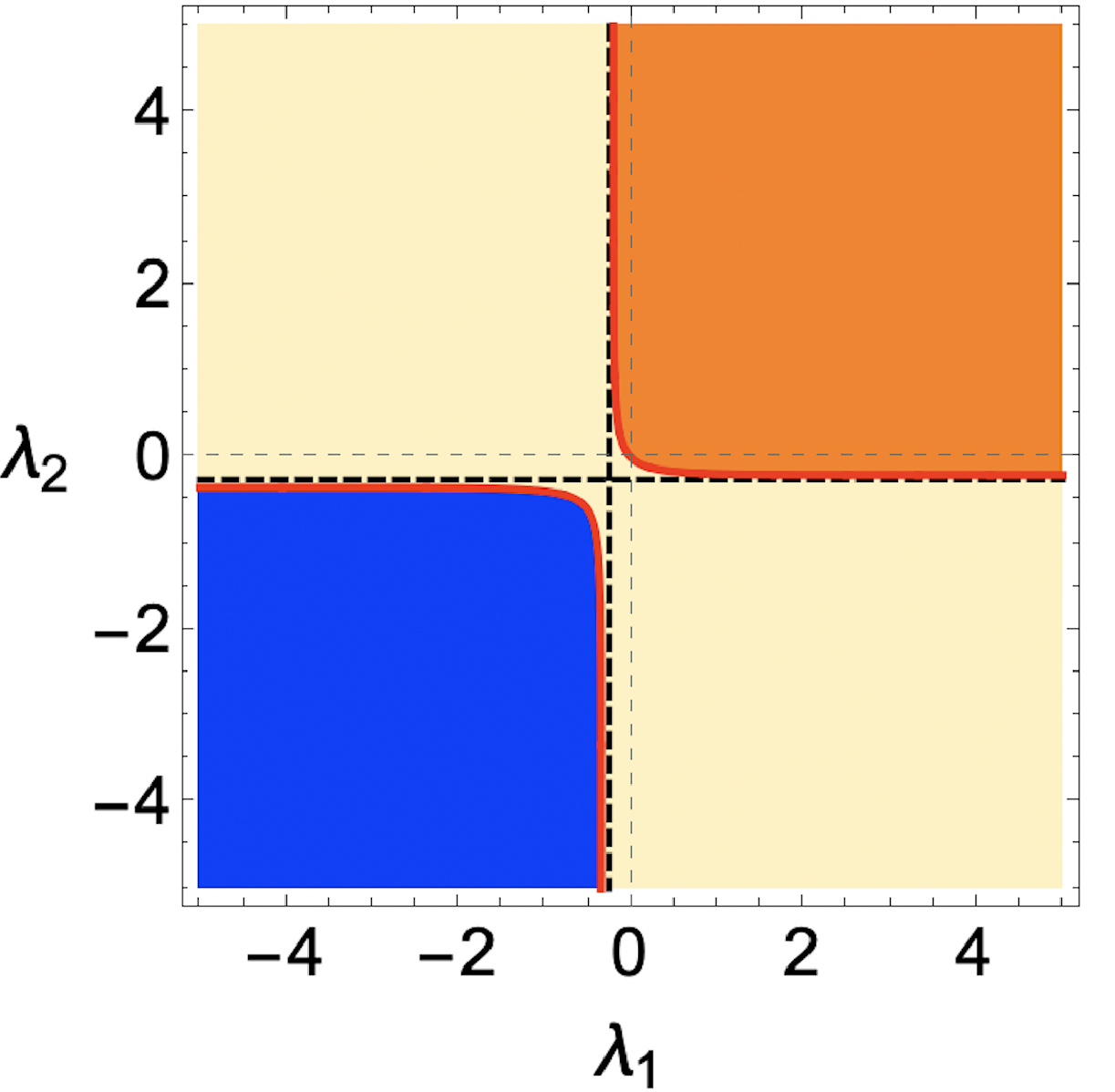} \qquad \qquad \qquad  \includegraphics[scale=0.32]{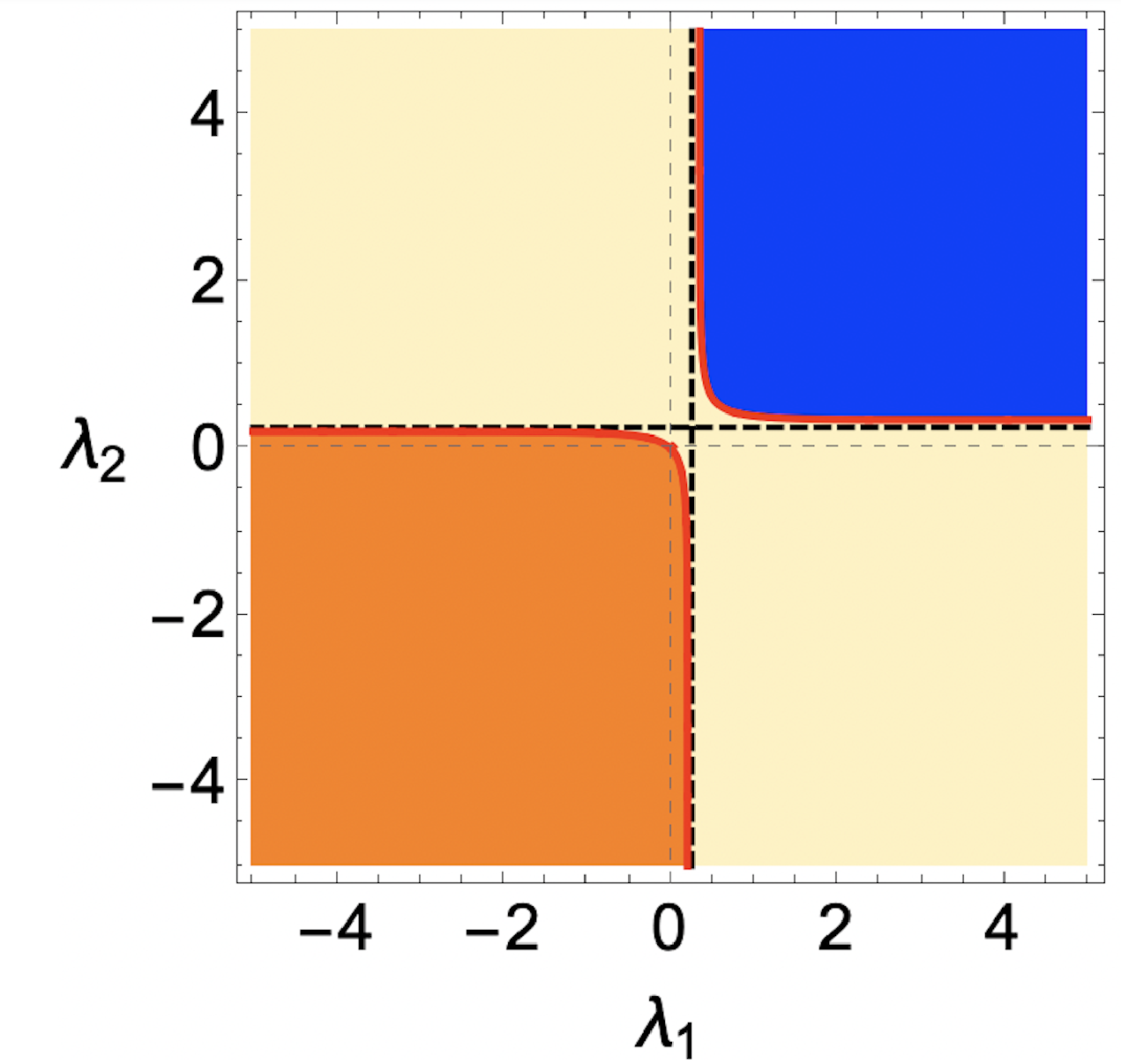}
\caption[\footnotesize Electron (left) and positron (right) bound state map for a double mass-spike contact interaction]{\footnotesize Electron (left) and positron (right) bound state map for a double mass-spike contact interaction. Blue area: 2 bound states. Yellow area: 1 bound state. Orange area: no bound states. The red line is the hyperbolic transcendent equation \eqref{idbtan} in the left plot and \eqref{idbtanp} in the right one. In this plot $a=m=1$.}
\label{fig:mapmasel}
\end{figure}
Contrary to what happens in the case of the  electric double delta potential, here there are no zero modes ($\omega=0$) in the spectrum because the critical points of the function in the right hand side of the equality \eqref{specc} occurs when $\kappa=-m\coth[(\lambda_1+\lambda_2)/2] $ or $\kappa=-m\tanh[(\lambda_1+\lambda_2)/2]$, and consequently $\lambda_1+\lambda_2$ should be infinite in order for $\kappa=m$ to hold.

\item  \underline{Positron bound state spinors: $0<\kappa<m$.}

An analogous procedure that the one applied for electrons yields  the following transcendent equation:
\begin{equation}
\coth \lambda_1 + \coth \lambda_2=\frac{4}{p}. \label{idbtanp}
\end{equation}
This hyperbolic transcendent equation also describes in the $\lambda_1$-$\lambda_2$ plane a curve similar to an ordinary hyperbola with two branches. One of the vertices is placed at the origin, the other at the point $(\lambda_1= \lambda_2={\rm arccoth}\, 2)$ in the first quadrant, and the axis is the $\lambda_1=\lambda_2$ straight line. For points above the upper branch of the curve, two bound states are encountered. Points in between the two branches correspond to one bound state. Points in the zone below the lower branch correspond to no bound states, as could be seen in Figure \ref{fig:mapmasel} (right). Once more, there are no zero modes.

\end{itemize}

\subsubsection{Electron and positron scattering spinors: the continuous spectrum}
\label{subsection4.2.2}
\begin{itemize}

\item \underline{Electron scattering spinorial waves: $k\in\mathbb{R}$}.

Replacing \eqref{eq22} and \eqref{eq23} in \eqref{matchdoble} for $q_1=q_2=0$ allows to obtain a pair of  algebraic lineal systems of four equations for the four unknowns of the ``{\it diestro}'' scattering $\{\sigma_R,A_R, B_R,\rho_R\}$, and for the four unknowns of the ``{\it zurdo}'' scattering $\{\sigma_L,A_L, B_L,\rho_L\}$. The solution for the electron scattering amplitudes reads:
\vspace{5pt}
\begin{eqnarray*}
&&\hspace{-0.6cm}\sigma_R (k;\lambda_1,\lambda_2)=\sigma_L(k;\lambda_1,\lambda_2)=\frac{k^2}{\Delta(k;\lambda_1,\lambda_2)}=\sigma(k;\lambda_1,\lambda_2), \nonumber\\
&&\hspace{-0.6cm}\rho_R(k;\lambda_1,\lambda_2)=\frac{-2im\sqrt{k^2+m^2}\sinh \lambda_1 \sinh \lambda_2 \sin(2ak)-ik \sqrt{k^2+m^2}\, \Upsilon(k;\lambda_1,\lambda_2)}{\Delta(k;\lambda_1,\lambda_2)},\nonumber\\
&&\hspace{-0.6cm}\rho_L(k;\lambda_1,\lambda_2)=\frac{-2im\sqrt{k^2+m^2}\sinh \lambda_1 \sinh \lambda_2 \sin(2ak)-ik \sqrt{k^2+m^2}\, \Upsilon^*(k;\lambda_1,\lambda_2)}{\Delta(k;\lambda_1,\lambda_2)},\nonumber\\ 
&&\hspace{-0.6cm}A_L(k;\lambda_1,\lambda_2)= B_R(k;\lambda_1,\lambda_2)=\frac{-ik\sqrt{k^2+m^2} \, e^{2iak} \sinh\lambda_1}{\Delta(k;\lambda_1,\lambda_2)},\nonumber\\
 &&\hspace{-0.6cm}B_L(k;\lambda_1,\lambda_2)=A_R(k;\lambda_1,\lambda_2)=\frac{k^2 \cosh \lambda_1+ i k m \sinh \lambda_1}{\Delta(k;\lambda_1,\lambda_2)},\nonumber\\
&&\hspace{-0.6cm}\Delta(k;\lambda_1,\lambda_2)=k^2 \cosh(\lambda_1\!+\!\lambda_2)+(k^2+m^2) (e^{4iak}-1)\sinh\lambda_1 \sinh\lambda_2 +i k m \sinh(\lambda_1\!+\!\lambda_2),\nonumber\\
&&\hspace{-0.6cm}\Upsilon(k;\lambda_1,\lambda_2) =e^{-2iak}\cosh \lambda_2 \sinh \lambda_1 + e^{2iak} \cosh \lambda_1 \sinh \lambda_2.
\label{elscatdosmas}
\end{eqnarray*}
Only if $\lambda_1=\lambda_2$ the scattering process is parity invariant, and $\rho_L=\rho_R$. If $k=i\kappa$ and $0<\kappa<m$, the zeroes of $\Delta(i \kappa;\lambda_1,\lambda_2)$, i.e. the poles of $\sigma(k;\lambda_1,\lambda_2)$ in the positive imaginary axis of the $k$-complex plane, enable to recover the zeroes of the transcendent equation \eqref{specc}.

\item \underline{Positron scattering spinorial waves: $k\in\mathbb{R}$}.

Proceeding in a similar way as explained before,  scattering of positrons by two mass-like $\delta$-impurities can be analysed obtaining the following scattering amplitudes:
\begin{eqnarray*}
&&\hspace{-0.6cm}\tilde{\sigma}_R(k;\lambda_1,\lambda_2)=\tilde{\sigma}_L(k;\lambda_1,\lambda_2)=\frac{k^2}{\tilde{\Delta}(k;\lambda_1,\lambda_2)}=\tilde{\sigma}(k;\lambda_1,\lambda_2),\nonumber\\
&&\hspace{-0.6cm}\tilde{\rho}_R(k;\lambda_1,\lambda_2)=\frac{i 2m \sqrt{k^2+m^2}\sinh \lambda_1 \sinh \lambda_2 \sin(ak) -ik\sqrt{k^2+m^2}\, \tilde{\Upsilon}(k;\lambda_2, \lambda_1)}{\tilde{\Delta}(k;\lambda_1,\lambda_2)},\nonumber\\
&&\hspace{-0.6cm}\tilde{\rho}_L(k;\lambda_1,\lambda_2)=\frac{i 2m \sqrt{k^2+m^2}\sinh \lambda_1 \sinh \lambda_2 \sin(ak) -ik\sqrt{k^2+m^2}\, \tilde{\Upsilon}^*(k;\lambda_2, \lambda_1)}{\tilde{\Delta}(k;\lambda_1,\lambda_2)},\nonumber\\
&&\hspace{-0.6cm}\tilde{A}_L(k;\lambda_1,\lambda_2)=\tilde{B}_R(k;\lambda_1,\lambda_2)= \frac{-i \, k \, \sqrt{k^2+m^2} e^{2iak} \sinh \lambda_1}{\tilde{\Delta}(k;\lambda_1,\lambda_2)},
\end{eqnarray*}
\begin{eqnarray*}
&&\hspace{-0.6cm}\tilde{B}_L(k;\lambda_1,\lambda_2)= \tilde{A}_R(k;\lambda_1,\lambda_2)=\frac{k^2 \cosh \lambda_1 - i \, k \, m \sinh \lambda_1}{\tilde{\Delta}(k;\lambda_1,\lambda_2)},\nonumber\\
&&\hspace{-0.6cm}\tilde{\Delta}(k;\lambda_1,\lambda_2)= k^2 \cosh(\lambda_1\!+\!\lambda_2)+(k^2+m^2) (e^{4iak}-1)\sinh\lambda_1 \sinh\lambda_2 - i k m \sinh(\lambda_1\!+\!\lambda_2),\nonumber\\
&&\hspace{-0.6cm}\tilde{\Upsilon}(k,\lambda_1, \lambda_2)= e^{-2iak}\cosh \lambda_2 \sinh \lambda_1+e^{2iak} \cosh \lambda_1 \sinh \lambda_2.
\label{posscatdosmas}
\end{eqnarray*}
Again, only whether $\lambda_1=\lambda_2=\lambda$ then $\tilde{\rho}_R(k;\lambda,\lambda)=\tilde{\rho}_L(k;\lambda,\lambda)$, and positron scattering through two $\delta$-impurities is parity invariant. 
\end{itemize}

\noindent All the $S$-matrices defined in this section both for electrons and positrons are unitary (i.e. $S^\dagger S = \mathbb{I}$), since it can be checked that 
\begin{eqnarray*}
|\sigma|^2 + |\rho_R|^2=1, \qquad |\sigma|^2 + |\rho_L|^2=1, \qquad \sigma \rho_L^* + \sigma^* \rho_R =0.
\end{eqnarray*}

\section{Second quantisation and vacuum energy at zero temperature}
\label{Sec5}
In order to build a relativistic QFT, one can postulate the operator
\begin{equation*}
\hat{\Psi}(t,z)= \int \frac{dk}{\sqrt{4\pi \omega}}  \left[ \hat{b}(k) u_+(k) e^{-i\omega t} e^{ikz}+ \hat{d}^{\dagger}(k) v_+(k) e^{i\omega t} e^{-ikz} \right], \qquad \textrm{with}\qquad \omega= +\sqrt{m^2+k^2},
\end{equation*}
which satisfies the Dirac equation, and interpret the coefficient $\hat{b}$ as a particle-annihilation  operator upon the second quantisation is performed. $\hat{d}$ would be an antiparticle-annihilation operator \cite{Taylorbook}. On the contrary, $\hat{b}^\dagger, \hat{d}^\dagger$ create nucleons and antinucleons of momentum $k$, respectively. When dealing with fermions, the corresponding states should be antisymmetric to enforce Pauli's exclusion principle, and hence $\hat{b}, \hat{d}$ fulfil the usual anti-commutation relations. The introduction of antiparticles enables to study the charge conjugation symmetry. As stated in \cite{GuilarteFrontiers2019}, for the specific choice of the Clifford algebra representation \eqref{gamma-mat}, the point supported potential \eqref{def-gen-pot} is invariant under  parity transformation defined by $\mathcal{P} \hat{\Psi}(t, z)\mathcal{P}^{-1}= \eta_p \gamma^0 \hat{\Psi}(t, -z)$, time reversal transformations $\mathcal{T} \hat{\Psi}(t, z)\mathcal{T} ^{-1}=\eta _T \gamma^0 \hat{\Psi}(-t, z),$ but not under charge conjugation $\mathcal{C} \hat{\Psi}(t, z)\mathcal{C} ^{-1}=\eta_C \gamma^2 \hat{\Psi}^*(t, z)$ as long $q\neq 0$ because $T_\delta^C(q,\lambda) = T_\delta(-q,\lambda)$. Furthermore, parity and time reversal are intrinsic symmetries of $H_\Psi^{(0)}$ and $H_\Phi^{(0)}$ for any choice of the Clifford algebra.  Charge conjugation is neither a symmetry of the Dirac Hamiltonian for fermions nor for antifermions since $\mathcal{C}H_\Psi^{(0)}\mathcal{C}^{-1}= H_\Phi^{(0)}$.  On the other hand, the CPT theorem \cite{Schwinger1951c, Luders1957, Pauli1957} ensures that every relativistic Quantum Field Theory is invariant under a simultaneous $\mathcal{CPT}$ transformation.  In this case it is clear that $\mathcal{CPT}H_\Psi^{(0)} (\mathcal{CPT})^{-1}= H_\Phi^{(0)}$. Since the space of states has been defined as the tensor product of the space of eigen-states of $H_\Psi^{(0)}$ times that of $H_\Phi^{(0)}$, the $\mathcal{CPT}$ symmetry  only permutes the order of the components in the tensor product. This swap keeps the total space unchanged, as was also the case for the charge conjugation symmetry.

Another interesting question in QFT is the study of the quantum vacuum energy. The problem of a Dirac field confined in a finite interval $[-L/2, L/2]$ has been studied in \cite{AIMIJG2015, DonaireSym2019, Alvarez2015}. The main idea explained in these works is that the Dirac Hamiltonian is not self-adjoint while restricted to square-integrable spinors defined in a finite interval, and there is a non zero flux of charge density through the boundaries. However, the Dirac Hamiltonian admits an infinite set of self-adjoint extensions in one-to-one correspondence with local unitary operators related to the boundary conditions. Hence, the domain of the self-adjoint extensions is the set of square-integrable spinors in $[-L/2, L/2]$ that satisfy some specific boundary conditions for their two components \cite{DonaireSym2019}. As a consequence, the unitarity of the QFT translates into a charge conservation for the Dirac field in the finite interval. It is interesting to note that the self-adjoint extension of the Dirac operator that represents the interaction of the quantum field with the boundary is given by the general M.I.T. bag boundary condition\footnote{The M.I.T. bag model was proposed by A. Chodos, R. L. Jaffe, K. Johnson, C.B. Thorn and V. Weisskopf \cite{Chodos1974} to study the confinement of quarks in the hadron model. This bag is a classical spherical cavity with quarks and gluons moving freely but confined inside it. In QFT, this implies that the fermionic current through the surface of the bag must be equal to zero and thus, certain specific boundary conditions must be imposed over the field representing the quarks, as also explained in \cite{ElizaldeIntJMPA2003}.  Due to the confinement, there are no exterior modes. This confinement gives rise to the Casimir effect (see \cite{ElizaldeJPA1998, Johnson1975, Milton1983} and references therein). }.

Whenever the fermionic field is confined between plates and hence, restricted to live in a finite interval, the whole spectrum of normal modes is discrete and it can be obtained from  the zeroes of a spectral function $h(k)$. In such a way, the quantum vacuum interaction energy can be computed by using the Cauchy's theorem of complex analysis as: 
\begin{eqnarray}
&&\hspace{-17pt}\! E_0= - \!\! \sum_{k\in \mathbb{R}^+}  2 \left. \sqrt{m^2+k^2} \right|_{h(k)=0} \!= \oint_C \frac{i \, dk}{\pi}\,  \sqrt{m^2+k^2} \, \, \partial_k \log h(k) ,
\label{DonaireE0fer}
\end{eqnarray}
where the minus sign in the first equality has been added due to the negative energy of the Dirac sea, and the factor 2 comes from the fact that positrons and electrons contribute in the same way to the summation over the spectrum. In \cite{DonaireSym2019} the contour $C$ is chosen as a semiring of inner radius $m$ and outer infinite radius satisfying $Re(k)>m$. This formalism explained in \cite{DonaireSym2019} can be also applied in the present work for two specific cases:
\begin{enumerate}
\item If $\lambda=0$ and $q=\pi$. It can be checked that in this case $T=-\id_2$. Following \cite{DonaireSym2019},  the spectral function is:
\begin{equation*}
h(k)=(-\sqrt{m^2+k^2} +m ) \sin (2ka).
\end{equation*}
From this point one could compute $E_0$ by using the equation \eqref{DonaireE0fer}, and by subtracting the divergences in a similar way that the one explained in  \cite{JMMCEPJC2020, BordagMPLA2020}. It can be checked in Figure 3 of \cite{DonaireSym2019} that for heavy fermions such that $ma=10$ the numerical result for the quantum vacuum interaction energy between plates is positive.
\item If $q_r= \sqrt{ \lambda^2+\pi^2 r^2}$, being $r \in \mathbb{Z}-\{0\}$. It is easy to show that now $T=-\id_2$ if $r$ is an odd number and $T=\id_2$ if $r$ is even. The resulting spectral functions take the form:
\begin{equation*}
h(k)=(\mp \sqrt{m^2+k^2}+m) \sin(2ka).
\end{equation*}
The quantum vacuum interaction energy between plates  is also positive in this occasion.
\end{enumerate}
These two cases are essentially analogous to one presented for bosons in \cite{JMMCEPJC2020}, when the boundary conditions mimicking the plates are Dirichlet or Neumann ones. In that particular instance, the plates became physically opaque and fluctuation propagation was restricted to the  compact space between plates.  However, when the boundary condition defined by $T(q, \lambda)$ in \eqref{Tmatrixbc} is not unitary, or if the fermionic field also lives outside the finite interval $[-L/2, L/2]$, the method explained in \cite{JMMCEPJC2020, BordagMPLA2020} and used in \cite{DonaireSym2019} can no longer be applied. It would be necessary to approach the problem from other perspectives that go beyond the limits of this work. 

\section{Conclusions}
\label{Sec6}
In this work I study the spectrum of bound and scattering states of relativistic fermionic particles interacting with double  Dirac $\delta$-potential in (1+1)-dimensional theories. The fermions propagating on the real line are interpreted as quanta emerging from the spinor fields. The problem has been addressed by solving at the same time the spectral problem of either the Dirac Hamiltonian $H_\Psi$ and its conjugate $H_\Phi$ in one-dimensional relativistic quantum mechanics. The eigen-spinors of both Hamiltonians have been interpreted as the one particle states with positive energy to be occupied by electrons and positrons after the fermionic second quantisation procedure be implemented. It has also been explained that the boundary condition matrix which represents the $\delta$-potential is parity and time-reversal invariant but it is not charge-conjugated invariant for the specific choice of the representation of the Clifford algebra $	\{\gamma^0=\sigma_3, \gamma^1=i\sigma_2, \gamma^2=\sigma_1\}$.

I deal with two particular cases, namely a double electric $\delta$-potential and a double mass-spike $\delta$-potential. In both cases,  the transcendent equations for computing the momenta of the bound states have been completely determined. It has been possible to elaborate a map of the number of bound states (two, one or zero) and zero modes present in the problem for a specific choice of $\{q_1, \, q_2, \lambda_1, \lambda_2, \, p^{-1}=am\}$. Notice that this map would be crucial to build the associated QFT in future works, because the states with negative energy will break the unitarity of the QFT and a mass term must be introduced to avoid the absorption phenomena problem. It has been shown that in the electric case, there could be zero, one  and two bound states as well as zero modes depending on the values of the parameters. Furthermore, the bi-parametric family of theories indexed by the coefficients of the $\delta$-functions is in one-to-one correspondence with a subset of the  moduli of complex tori. The topology of the torus is determined by the two angular coordinates given by $q_1, q_2$. The complex structure of the torus is completely characterised by $p^{-1}$, i.e. by the product of the mass of the particles  and the distance between the two electric $\delta$-potentials. On the other hand, for the massive Dirac deltas case, there could be one, two or zero bound states depending of the value of $\lambda_1, \lambda_2$ and $p^{-1}$, but there are no zero modes. It is also worth stressing that either in the electric and the massive case, the $S$-matrices and the scattering data have been computed. 

Lastly, it has been completely understood that only if $\lambda=0, q=\pi$ or $q_r= \sqrt{ \lambda^2+\pi^2 r^2}$ with $ r \in \mathbb{Z}-\{0\}$, the boundary condition matrix $T_\delta(q, \lambda)$ which defines the self-adjoint extension of the Dirac Hamiltonian is unitary. In these cases, which represent totally opaque plates, the formalism developed in \cite{DonaireSym2019} can be applied  to compute the spectral function and the quantum vacuum interaction energy for fermions confined between plates.

The new objective for future work is to extend this work to the computation of  the Casimir force for a system of fermions under the influence of two general Dirac delta potentials. And whatsoever, once this system was studied, it would be straightforward to generalise it to fermions propagating in general Dirac delta-type lattices. This is a relevant topic in Condensed Matter Physics due to the edge states which appear in some meta materials that can be mimicked by these type of theories.  On the other hand, the analysis of the Green's function for fermions confined between plates modelled by general delta potentials in higher spatial dimensions will be left for further investigation.

\section{Acknowledgements}
This research was supported by Spanish MCIN with funding from European Union NextGenerationEU (PRTRC17.I1) and Consejer\'ia de Educaci\'on from JCyL through QCAYLE project, as well as MCIN project PID2020-113406GB-I00.  I am greatful to the Spanish Government for the FPU-PhD fellowship program (Grant No. FPU18/00957). I thank J.M. Guilarte and J.M. Mu\~{n}oz-Casta\~{n}eda for the interesting discussions on this subject.

\appendix
\section{Dirac double electric potential with $\boldsymbol{q_1=q_2=\pi/2}$, $\boldsymbol{m=1.5, a=1}$}

In the main part of this work, the general analytic formulas and data that characterise the bound and scattering states for fermions propagating under the influence of Dirac delta potentials have been collected, without going into particular cases. These general results will be of direct use in future work since they will appear in the calculations of the Casimir energy \cite{KK2008}, the density of states, and other thermodynamic quantities in the associated QFT.

Take into account that in this article the problem of positrons and electrons are treated separately, as two different problems. However, one can consider the case in which there are one bound state at the same time in each one of the two separate problems. In this appendix such a case is going to be considered. From now on, consider $q_1=q_2=\pi/2, m=1.5, a=1$.

\begin{figure}[H]
\centering
\includegraphics[scale=0.7]{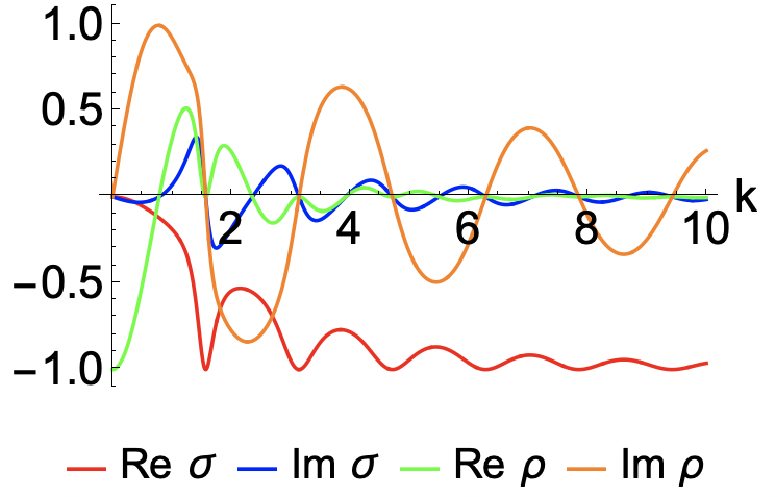} \quad \quad \quad 
\caption{\footnotesize Scattering coefficients $\sigma, \rho_R=\rho$ as a function of $k$ for the case $q_1=q_2=\pi/2, m=1.5, a=1$. }
\label{fig:8}
\end{figure}

\begin{figure}[H]
\centering
\includegraphics[scale=0.65]{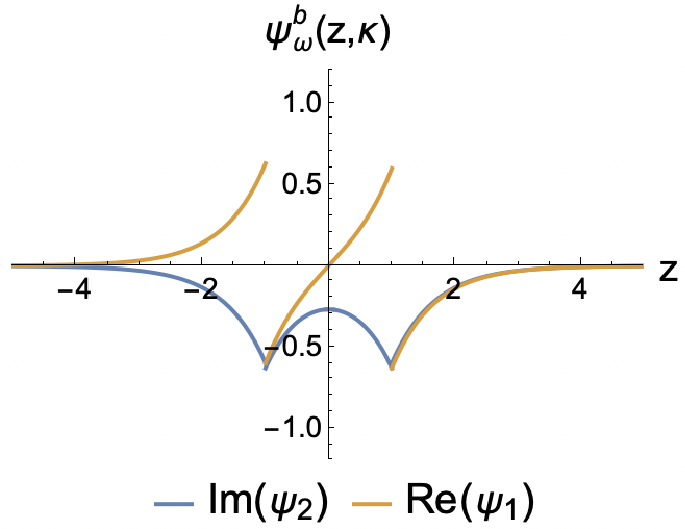}\quad \quad \includegraphics[scale=0.65]{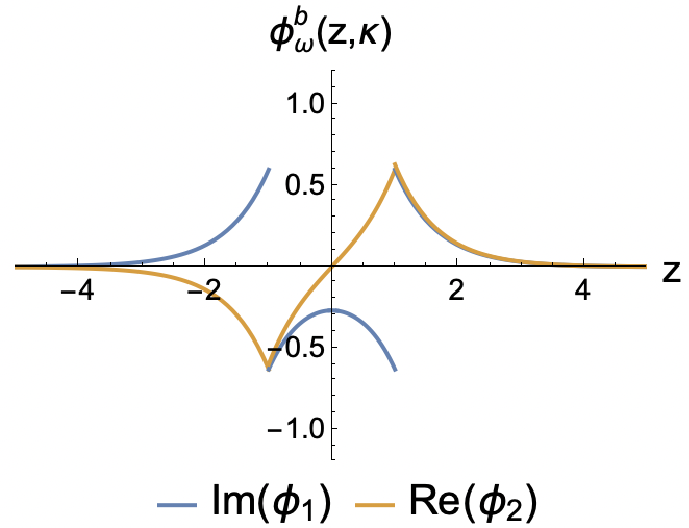}\quad \quad 
\caption{\footnotesize Left: Two non-zero spinor components for electrons as a function of $z$ for the particular case $q_1=q_2=\pi/2, m=1.5, a=1$. The numerical coefficients for this example take the following values: $A_1=1, B_2=-C_2=0.05004, D_3=-1, \, \mathcal{N}= \sqrt{7.92853}$. Right: Two non-zero spinor components for positrons as a function of $z$ for the case $q_1=q_2=\pi/2, m=1.5, a=1$. The numerical coefficients for this example take the following values: $A_1=1, B_2=C_2=-0.05004, D_3=1, \, \mathcal{N}= \sqrt{7.9285}$.}
\label{fig:9}
\end{figure}

Concerning the scattering, one could prove from the general equations in section \ref{subsection4.1.2} that $\rho_R=\rho_L=-\tilde{\rho}_R=-\tilde{\rho}_L$ and $\sigma=\tilde{\sigma}$. Both problems for electrons and positrons are thus time reversal and parity invariant. As can be seen in Figure \ref{fig:8}, all the scattering data is highly oscillatory for small values of the momentum. This behaviour is attenuated as $k$ increases. The value of the modulus of the transmission coefficient tends to one when the particle has high energy, as expected\footnote{In scattering theory, particles with high energy do not sense the potential and pass through it without been scattered.}. Concerning the bound state, the momentum solution of the trascendental equation \eqref{eq35} will be $\kappa=1.49813$. The corresponding wave function is represented in Figure \ref{fig:9}. The lower component of the electron spinor is continuous but not differentiable at the impurities. The electron/positron is shared by the two delta potentials. The upper component presents finite discontinuities at the points where the deltas are placed. There exists parity symmetry and wave functions are either even or odd linear combinations of the localised states of the fermions around each impurity. The spinor for the positron case is analogous. 

One could also consider other possible cases (two bound states for electrons, two for positrons, one bound state for electrons, one for positrons, no bound states in electrons and no in positrons, for both the massive and electric cases, as well as the case in which both electron and positron have a bound state at the same time in the massive Dirac delta case). The corresponding plots of the wave functions may change qualitatively but will maintain the same fundamental behavior described above, i.e., the real and imaginary parts of the two spinor components have either finite discontinuities at the points where the deltas are placed or have abrupt peaks that make the wave function continuous but not derivable at the same points. Furthermore, if the  modulus of the coefficients $q_1,q_2$ (or similarly $\lambda_1, \lambda_2$ in the double masive Dirac delta case) are not equal to each other, the wave functions may not have well-defined parity properties about the origin.

\section{Scattering data for double Dirac delta potentials. Specific examples}
In this appendix, the transmission and reflection coefficients will be analysed for the possible specific situations of the spectrum described in the article (two, one or zero bound states) in different configurations of both the electric and the massive double Dirac delta potential. Only the electron scattering coefficients are going to be considered in figures from \ref{fig:10} to \ref{fig:15}. The positron scattering data can be obtained from the general data in subsections \ref{subsection4.1.2} and \ref{subsection4.2.2} in a similar way.
\begin{itemize}
\item Case: $\lambda_1=\lambda_2=2, q_1=q_2=0, m=a=1$. The spectrum has no bound states. 
\begin{figure}[H]
\centering
\includegraphics[scale=0.5]{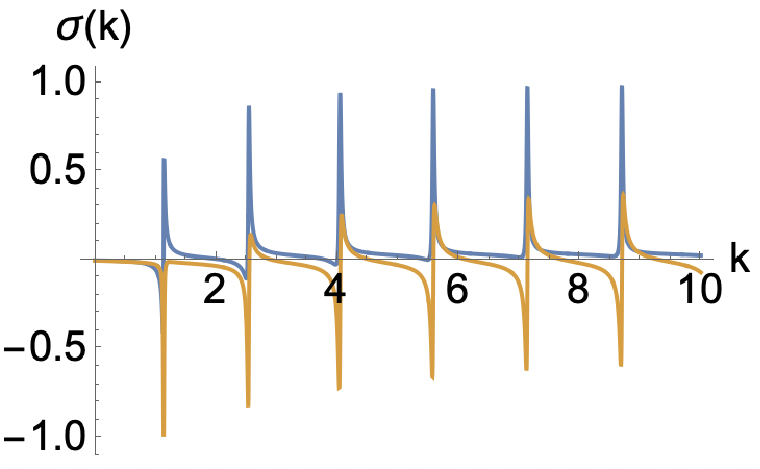} \quad \quad \quad \includegraphics[scale=0.5]{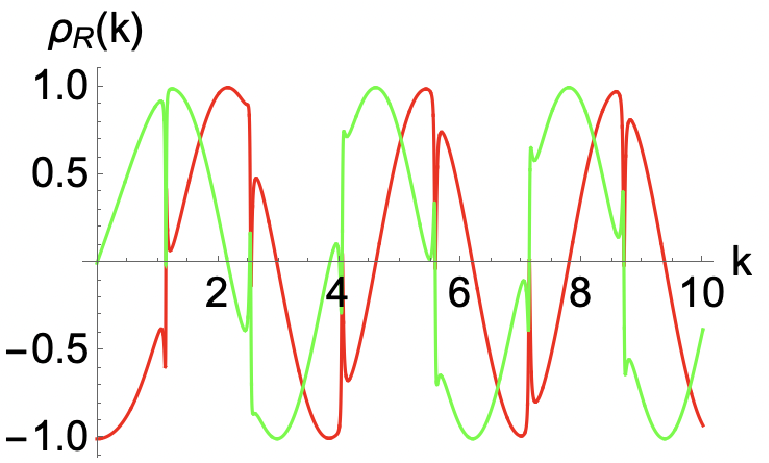}
\caption{\footnotesize Left: Real (blue) and imaginary part (orange) of $\sigma$ as a function of $k$. Right: Real (red) and imaginary part (green) of $\rho_R$ as a function of $k$. In both plots, $\lambda_1=\lambda_2=2, m=a=1$. }
\label{fig:10}
\end{figure}
\item Case: $\lambda_1=1, \lambda_2=-2, q_1=q_2=0, m=a=1$. The spectrum has one bound state. 
\begin{figure}[H]
\centering
\includegraphics[scale=0.5]{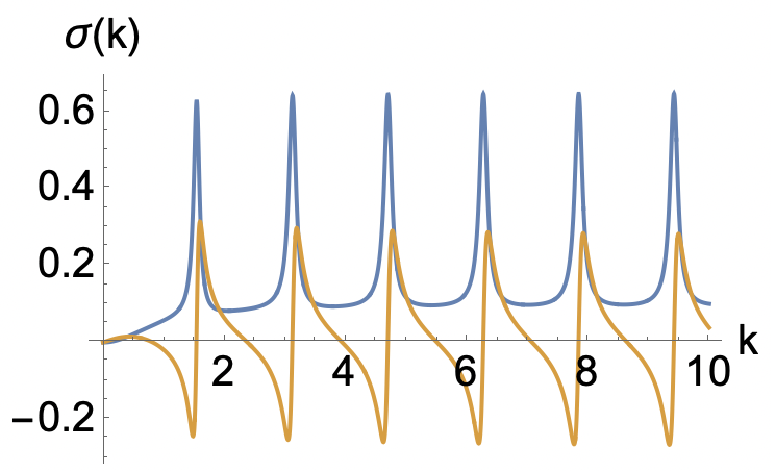} \quad \quad \quad \includegraphics[scale=0.5]{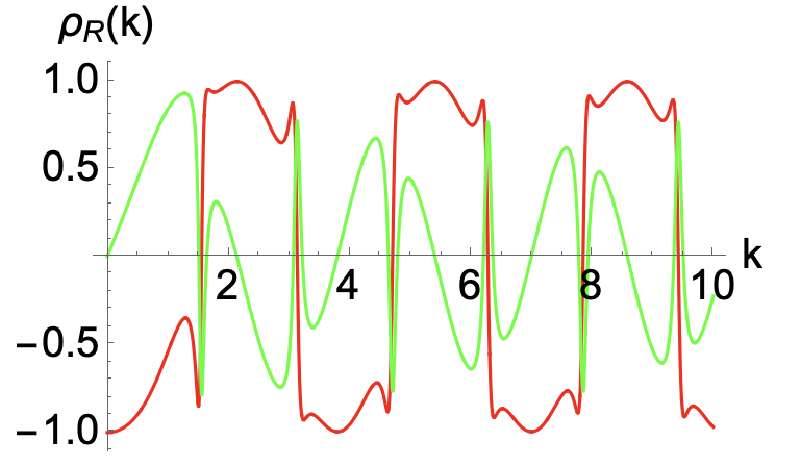}
\caption{\footnotesize Left: Real (blue) and imaginary part (orange) of $\sigma$ as a function of $k$. Right: Real (red) and imaginary part (green) of $\rho_R$ as a function of $k$. In both plots, $\lambda_1=1, \lambda_2=-2,  m=a=1$. }
\label{fig:11}
\end{figure}
\item Case: $\lambda_1=\lambda_2=-2, q_1=q_2=0, m=a=1$. The spectrum has two bound states. 
\begin{figure}[H]
\centering
\includegraphics[scale=0.5]{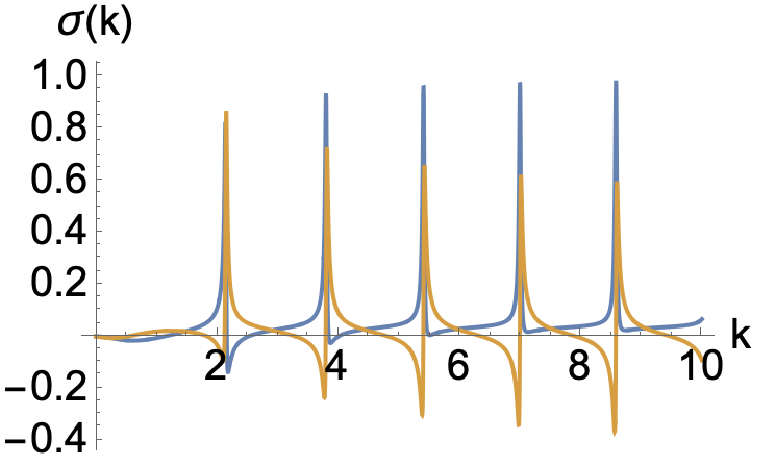} \quad \quad \quad \includegraphics[scale=0.5]{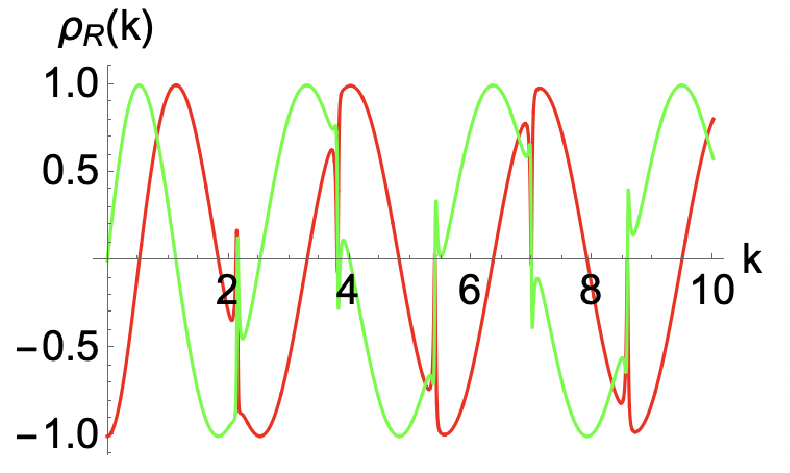}
\caption{\footnotesize Left: Real (blue) and imaginary part (orange) of $\sigma$ as a function of $k$. Right: Real (red) and imaginary part (green) of $\rho_R$ as a function of $k$. In both plots, $\lambda_1=\lambda_2=-2, m=a=1$. }
\label{fig:12}
\end{figure}
\item Case: $\lambda_1=\lambda_2=0, q_1=q_2=2, m=1.5, a=1$. The spectrum has two bound states. 
\begin{figure}[H]
\centering
\includegraphics[scale=0.5]{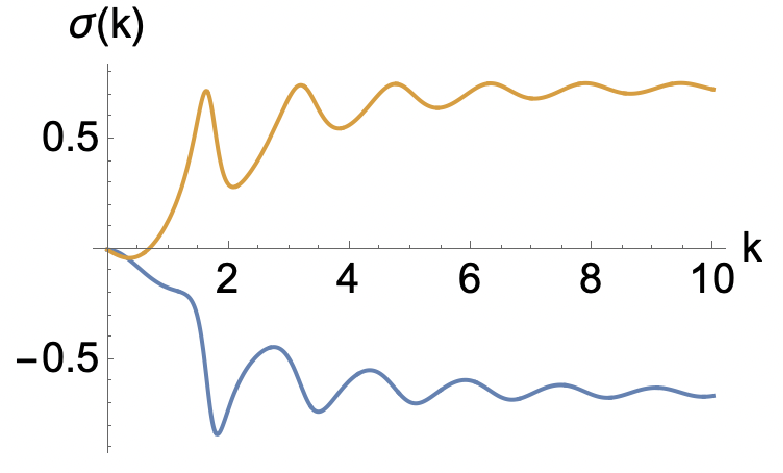} \quad \quad \quad \includegraphics[scale=0.5]{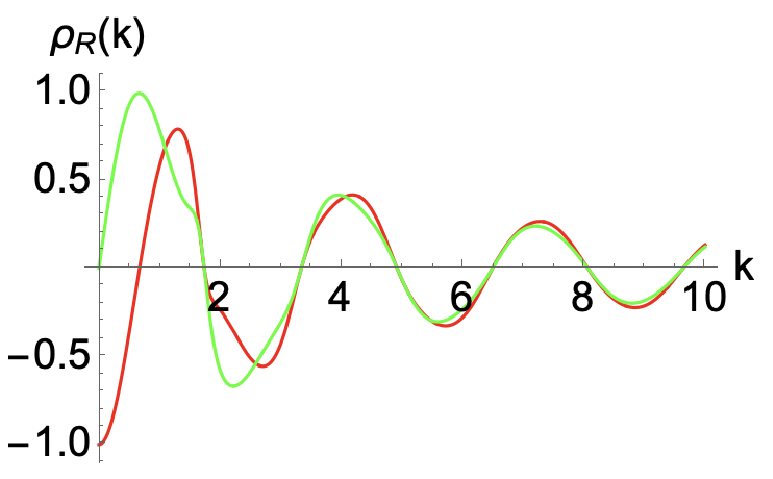} 
\caption{\footnotesize Left: Real (blue) and imaginary part (orange) of $\sigma$ as a function of $k$. Right: Real (red) and imaginary part (green) of $\rho_R$ as a function of $k$. In both plots, $q_1=q_2=2, m=1.5, a=1$. }
\label{fig:13}
\end{figure}
\item Case: $\lambda_1=\lambda_2=0, q_1=1, q_2=2, m=1.5, a=1$. The spectrum has one bound states.
\begin{figure}[H]
\centering
\includegraphics[scale=0.5]{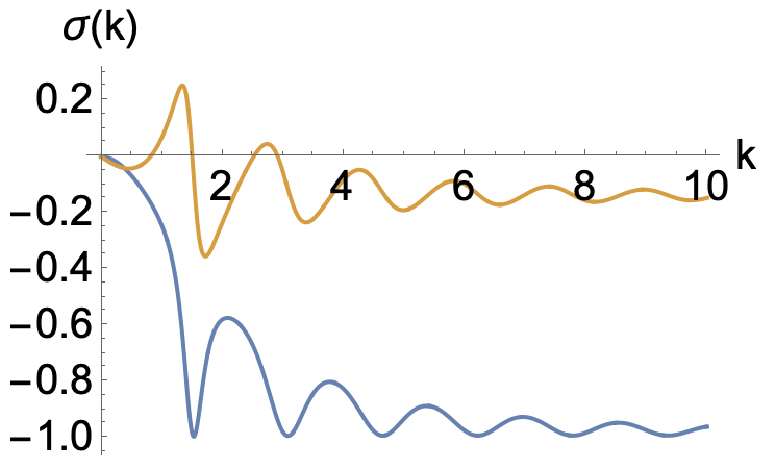} \quad \quad \quad \includegraphics[scale=0.5]{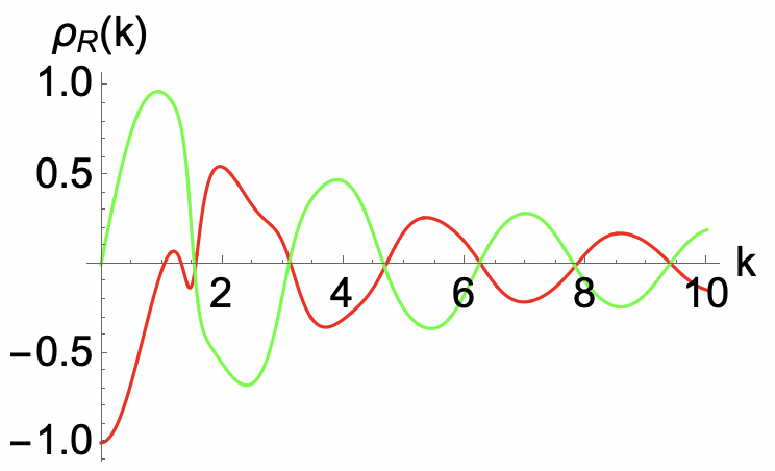} 
\caption{\footnotesize Left: Real (blue) and imaginary part (orange) of $\sigma$ as a function of $k$. Right: Real (red) and imaginary part (green) of $\rho_R$ as a function of $k$. In both plots, $q_1=1,q_2=2, m=1.5, a=1$. }
\label{fig:14}
\end{figure}
\item Case: $\lambda_1=\lambda_2=0, q_1=q_2=1, m=1.5, a=1$. The spectrum has no bound states. 
\begin{figure}[H]
\centering
\includegraphics[scale=0.5]{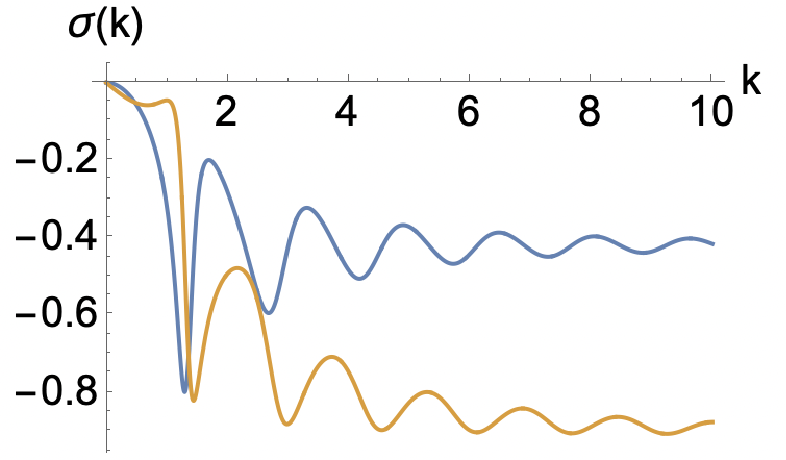} \quad \quad \quad \includegraphics[scale=0.5]{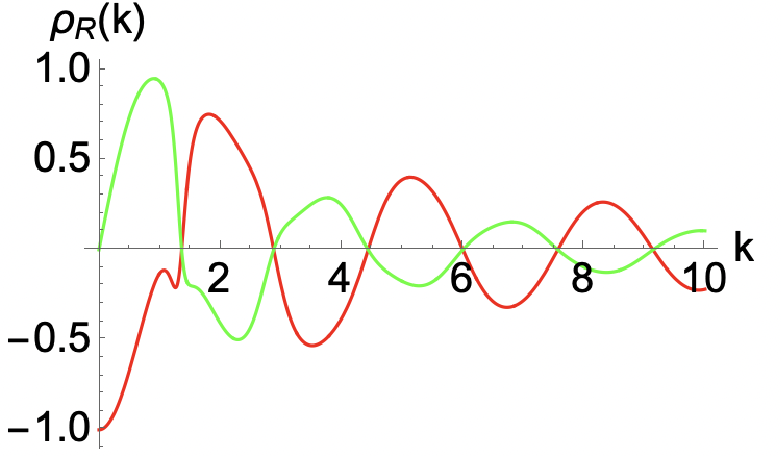}
\caption{\footnotesize Left: Real (blue) and imaginary part (orange) of $\sigma$ as a function of $k$. Right: Real (red) and imaginary part (green) of $\rho_R$ as a function of $k$. In both plots, $q_1=q_2=1, m=1.5, a=1$. }
\label{fig:15}
\end{figure}
\end{itemize}
At this point, some caveats are worth stressing. For the electric case the boundary matrix $T_\delta(q,0)$ in \eqref{Tmatrixbc} is a unitary matrix and, as can be seen from the plots, the scattering coefficients are such that $|\sigma|\to 1$ and $|\rho| \to 0$ as $k$ becomes large enough. This is in agreement with Chapter 4 of \cite{Galindobook}. On the other hand, for the massive case (with the exception of the choice\footnote{In this particular case, the matching condition $T_\delta(0, \lambda)$ is unitary, and $\sigma=1, \rho=0, \forall k.$} $\lambda=i \pi + 2i \pi n$ or $\lambda= 2i \pi n$ for $n \in \mathbb{R}$), the matrix $T_\delta(0, \lambda)$ is not unitary. Furthermore, it is easy to see from the plots above that some resonances appear for the scattering coefficients at some particular values of $k$. This situation can be checked to hold almost periodically as $k$ becomes large (a detailed study shows that the periodic interval is not exact, it decreases as $k$ increases). This resonances appear because in the spectrum of the quantum vacuum fluctuations there are also resonant states of the form $k=x_0+iy_0$ with $x_0>0$ and $y_0<0 \wedge |y_0|\ll1$ (see Figure \ref{fig:16}). They are poles of the unitary S-matrix. Moreover, it can be  seen in the figures above that in this case, the widths of the resonances do not increase significantly with energy and do not become progressively less pronounced above the background. This is slightly different from the case of a single delta potential explained in \cite{Galindobook}.
\begin{figure}[H]
\centering
\includegraphics[scale=0.25]{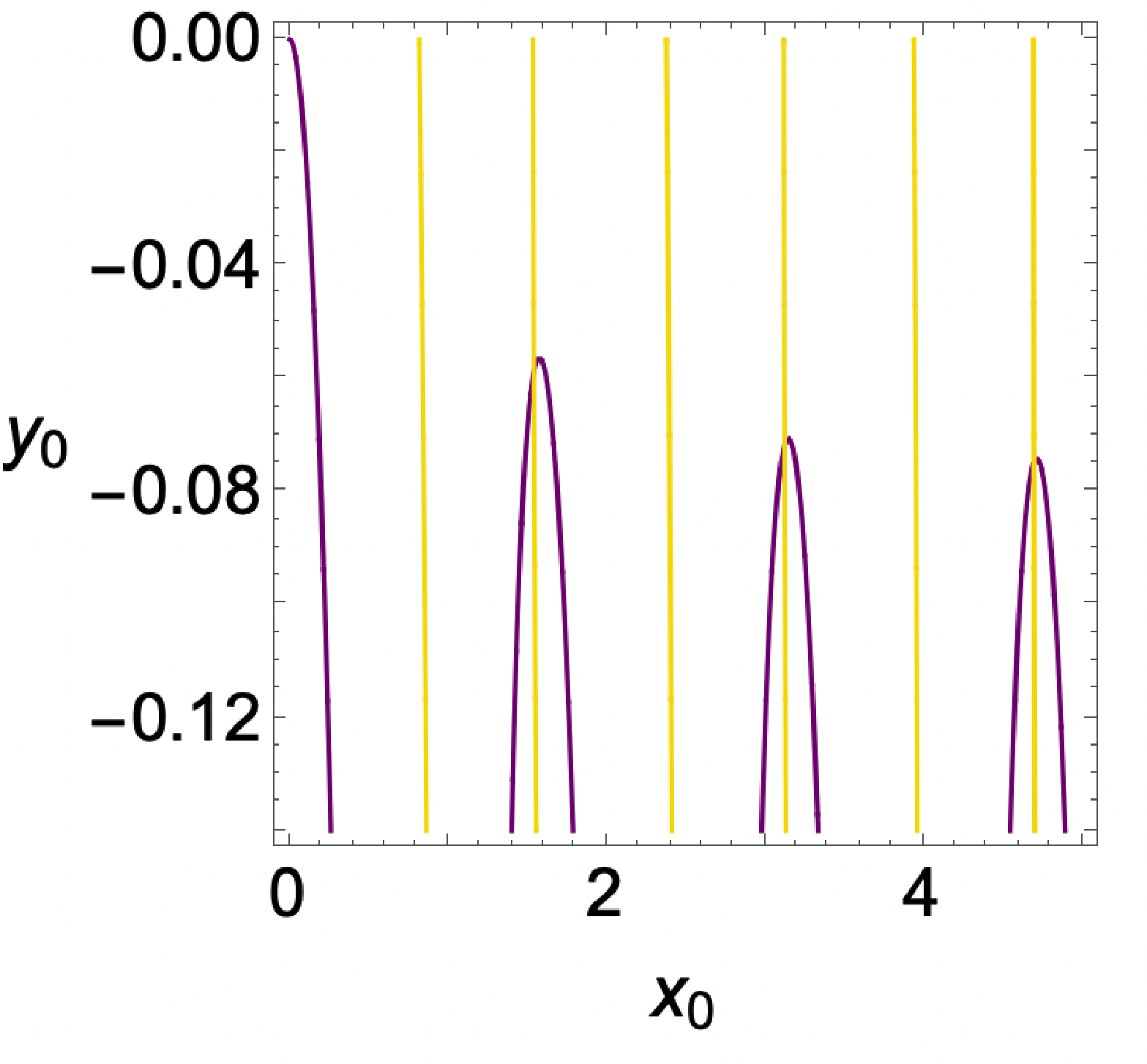} \quad \quad \quad 
\caption{\footnotesize Solutions of $\textrm{Re(Denominator of}\, \sigma)=0$ (purple) and solutions of $\textrm{Im(Denominator of}\, \sigma)=0$ (yellow) for the case $q_1=q_2=0, m=a=1, \lambda_1=1, \lambda_2=-2$. The intersecting points between both curves are poles of the S-matrix with the form $x_0+iy_0$ (i.e. they correspond to resonant states in the spectrum). }
\label{fig:16}
\end{figure}

\bibliographystyle{unsrt}
\bibliography{biblio}{}

\end{document}